\documentclass[aps,pre,epsf,superscriptaddress,amsmath,amssymb,amsfonts,twocolumn,showpacs]{revtex4-1}

\usepackage{graphicx}
\usepackage{epsfig}
\usepackage{dcolumn}
\usepackage{bm}
\usepackage{braket}
\usepackage{amsmath}
\usepackage{color} 
\newcommand{\abs}[1]{\left| #1 \right|}

\usepackage{tcolorbox}
\usepackage{tabularx}
\usepackage{array}
\usepackage{colortbl}
\tcbuselibrary{skins}
\usepackage{soul}

\newcolumntype{Y}{>{\raggedleft\arraybackslash}X}

\tcbset{tab1/.style={fonttitle=\bfseries\large,fontupper=\normalsize\sffamily,
colback=yellow!10!white,colframe=red!75!black,colbacktitle=white!40!white,
coltitle=black,center title,freelance,frame code={
\foreach \n in {north east,north west,south east,south west}
{\path [fill=red!75!black] (interior.\n) circle (3mm); };},}}

\tcbset{tab2/.style={enhanced,fonttitle=\bfseries,fontupper=\normalsize\sffamily,
colback=yellow!6!white,colframe=red!50!black,colbacktitle=yellow!20!white,
coltitle=black,center title}}

\begin{document}
\title{Pump Probe Spectroscopy of Bose Polarons:\\ Dynamical Formation and Coherence}

\author{S.I. Mistakidis}
\affiliation{Center for Optical Quantum Technologies, Department of Physics, University of Hamburg, 
Luruper Chaussee 149, 22761 Hamburg Germany}

\author{G.C. Katsimiga}
\affiliation{Center for Optical Quantum Technologies, Department of Physics, University of Hamburg, 
Luruper Chaussee 149, 22761 Hamburg Germany}

\author{G.M. Koutentakis}
\affiliation{Center for Optical Quantum Technologies, Department of Physics, University of Hamburg, 
Luruper Chaussee 149, 22761 Hamburg Germany}\affiliation{The Hamburg Center for Ultrafast Imaging,
University of Hamburg, Luruper Chaussee 149, 22761 Hamburg, Germany}

\author{Th. Busch}
\affiliation{OIST Graduate University, Onna, Okinawa 904-0495, Japan}

\author{P. Schmelcher}
\affiliation{Center for Optical Quantum Technologies, Department of Physics, University of Hamburg, 
Luruper Chaussee 149, 22761 Hamburg Germany} \affiliation{The Hamburg Center for Ultrafast Imaging,
University of Hamburg, Luruper Chaussee 149, 22761 Hamburg,
Germany}

\date{\today}

\begin{abstract} 
We propose and investigate a pump-probe spectroscopy scheme to unveil the time-resolved dynamics 
of fermionic or bosonic impurities immersed in a harmonically trapped Bose-Einstein 
condensate.
In this scheme a pump pulse initially transfers 
the impurities from a noninteracting to a resonantly interacting spin-state and, after a finite time in which the system evolves freely,
the probe pulse reverses this transition. 
This directly allows to monitor the nonequilibrium dynamics of the impurities as
the dynamical formation of coherent attractive or repulsive Bose polarons and signatures of their induced-interactions are 
imprinted in the probe spectra. 
We show that for interspecies repulsions exceeding the intraspecies ones a temporal orthogonality catastrophe occurs, followed 
by enhanced energy redistribution processes, independently of the impurity's flavor. 
This phenomenon takes place over the characteristic trap timescales. 
For much longer timescales a steady state is reached characterized by substantial losses of coherence of the impurities.   
This steady state is related to eigenstate thermalization and it is demonstrated to be independent of the system's characteristics. 
\end{abstract}

\maketitle

\section{Introduction} 

Time-resolved spectroscopy is an established technique for the 
characterization of the dynamical response of a wide range of physical systems~\cite{Demtroder}. 
The general idea underlying a pump-probe spectroscopy (PPS) scheme is that a pump pulse prepares a nonstationary state 
of the system under consideration, which is then {\it interrogated} by a time-delayed probe pulse. 
This allows for simultaneous spectral and temporal resolution of the induced dynamical processes, 
exposing the energy redistribution of the selectively triggered excitations \cite{Disa,Lara_Astiaso}, 
in sharp contrast to time-independent 
spectroscopic techniques like injection spectroscopy~\cite{Levenson,Kohstall,Koschorreck,Cetina}. 
Applications of the PPS protocol range from two- and three-level atomic 
systems~\cite{Mollow,Berman,Wei,Wu,Mills,Lukin,Li} to the ultrafast dynamics of 
photoexcited quantum materials~\cite{Stock,Wu1,Sotier,Eisele,Takei,Kolarczik,Caval1,Caval2}. 
Such a time-domain analysis has been proven to be a powerful tool for resolving the ultrafast 
molecular dynamics allowing for instance for a coherent control of bound excited-state dimers over 
long timescales~\cite{Machholm,McCabe}. 
PPS has also been utilized for studying the pair-correlation dynamics of ultracold Bose gases~
\cite{Koch}, offering a potential connection between ultrafast and ultracold physics~\cite{McCabe,Salzmann}. 

Operating in the ultracold regime, in this work we propose a PPS scheme as a toolkit for 
investigating in a time-resolved manner the {\it impurity problem} and the related 
formation of and interactions between quasiparticles~\cite{Catani,Fukuhara,Yan_pol,Scelle,Schmidtred,Ardila2, 
Artemis2,Guenther,Mayer,Jorgensen,Hu,Shchadilova,Levinsen,Christensen,Grusdt2,Mistakidis_bose_pol,
Enss_beyond,Us_Busch,us,Sdiss,Bipolarons_Camacho,eff_int_Camacho,res_int_Ardila,Schirotzek,Navon,
Chevy,Pilati,Massignan1,Schmidt2,Massignan2,Massignan3,
Burovski,Scazza,Schmidt3,Burovski,Gamayun,Us_Fermi,Fermi_catastr,Rydberg_imp_fermi}.
Understanding the physics of quasiparticles is important beyond cold atom settings in semiconducting~\cite{Semi} and superconducting devices~\cite{Ruggiero}. 
Additionally, interactions among quasiparticles in liquid Helium mixtures~\cite{He1,He2} and cuprates~\cite{Htc,Sous_bipolarons} are considered to be responsible for conventional and high-$T_c$ superconductivity~\cite{Cooper,Schrieffer,Alexandrov,Mott,Alexandrov1,Salje,Berciu}. 
Here we consider a Bose-Einstein condensate (BEC) with one or two impurities of either bosonic or fermionic nature immersed into it and track 
the emergent Bose polaron formation \cite{Catani,Fukuhara,Scelle,Schmidtred,Ardila2,Artemis2,
Guenther,Mayer,Jorgensen,Hu,Shchadilova,Levinsen,Christensen,Grusdt2,Mistakidis_bose_pol,Yan_pol,us,Enss_beyond,Us_Busch,Sdiss}  
with a PPS radiofrequency protocol analogous to the one used in the experiment of Ref.~\cite{Nobeler}. 
This allows us to probe and control the coherence properties of the quasiparticles. 
Our results pave the way for transferring the knowledge regarding the ultrafast dynamics 
of condensed matter systems~\cite{pola1,pola2,Orenstein,Giannetti} to the ultracold atomic realm. 

In our investigation, an intense pump pulse transfers the initially free bosonic or fermionic impurities
to an attractively or repulsively interacting state with the environment.  
After a variable dark time, during which the system evolves freely, a probe pulse of weaker intensity is applied, 
which de-excites the impurities. 
As the formation of well-defined attractive and repulsive Bose polarons in this many-body (MB) system is captured in 
the probe spectrum, this process allows to monitor the dynamics.
In systems where the interaction strength between the impurity and the background is not larger than the interaction 
strength within the background gas, polaronic excitations can have long lifetimes. 
However, beyond that limit substantial losses of coherence occur with a temporal orthogonality 
catastrophe (TOC)~\cite{Us_Busch,us,Sdiss,Goold} being imprinted in the probe spectrum. 
The TOC emerges due to the relaxation of the quasiparticles into energetically lower-lying, phase separated states. 
This process is independent of the number of the impurities or their statistics. 
Remarkably, for timescales longer than the characteristic confinement one, the probe spectrum unveils 
evidence towards eigenstate thermalization~\cite{Rigol1,Rigol_break_therm,Rigol2,Jansen}, 
where the impurities reside in an incoherent state characterized by a large effective temperature. 
This relaxation dynamics~\cite{Lausch1,Lausch2} is found to be independent 
of the size of the bath, the number and nature of the impurities, and their interaction strengths 
and mass. 

Our work is structured as follows. 
Section~\ref{model} introduces the setup under consideration and briefly comments on the employed variational 
approach to tackle the nonequilibrium dynamics of Bose polarons. 
In Sec.~\ref{PPS_details} we discuss the utilized PPS scheme and demonstrate the resulting Bose polaron 
spectrum for short and long evolution times with a particular focus on the impurity-impurity induced 
interactions, coherence properties and thermalization processes. 
In  Sec.~\ref{sec:interaction_energy} we elaborate 
on the emergent energy redistribution processes, while in order to gain further insights into the spectroscopically 
observed relaxation dynamics we invoke in Sec.~\ref{relaxation_ETH} 
the Eigenstate Thermalization Hypothesis (ETH). 
We summarize our results and provide an outlook including future perspectives in Sec.~\ref{conclusions}. 
Appendix~\ref{rf_implement} presents in detail the used radiofrequency spectroscopy scheme and Appendix~\ref{Ramsey} 
explicates briefly the predictions of a Ramsey protocol for strong impurity-medium interactions. 
The dimensional reduction of our MB Hamiltonian from three- to one-dimension is showcased in Appendix~\ref{dimensional_reduction}. 
Finally, Appendix~\ref{sec:method} deals with the variational method employed herein so as to simulate the PPS protocol 
and Appendix~\ref{convergence} delineates the convergence of the presented results. 

\section{Model Setup}\label{model}

Our model is a highly particle imbalanced  
mixture. It consists of   
$N_I=1,2$ bosonic or fermionic impurities (I) having 
a spin-$1/2$ degree of freedom~\cite{Kasamatsu} being immersed 
in a bosonic bath of $N_B=100$ structureless bosons (B). 
The mixture is assumed to be mass balanced, 
$m_B=m_I\equiv m$ (unless stated otherwise), while both species are harmonically 
confined in the same one-dimensional potential. 
Details of the dimensional reduction of our system are discussed in 
Appendix~\ref{dimensional_reduction}. 
The MB Hamiltonian reads
\begin{equation}
\hat{H} = \hat{H}^{0}_{B}+\hat{H}_{BB}+\textstyle\sum\limits_{a=\uparrow, \downarrow} (\hat{H}^{0}_a+ \hat{H}_{aa})
+\hat{H}_{\uparrow \downarrow}+\hat{H}_{BI}+\hat{H}_S^{\beta}.
\label{Htot_system}
\end{equation} 
Here, $\hat{H}^{0}_{B}=\int dx~\hat{\Psi}^{\dagger}_{B} (x) \left( -\frac{\hbar^2}{2 m_B} 
\frac{d^2}{dx^2} +\frac{1}{2} m_B \omega^2 x^2 \right) \hat{\Psi}_{B}(x)$, 
and $\hat{H}^{0}_a=\int dx~\hat{\Psi}^{\dagger}_a(x) \left( -\frac{\hbar^2}{2 m_I} \frac{d^2}{dx^2}  
+\frac{1}{2} m_I \omega^2 x^2 \right) \hat{\Psi}_a(x)$ denote the noninteracting 
Hamiltonian of the BEC and the impurities respectively 
while $a\in \left\{ \uparrow, \downarrow \right \}$.
Additionally, $\hat{\Psi}_{B} (x)$ [$\hat{\Psi}_{a} (x)$] 
is the field-operator of the BEC [spin-$a$ impurities]. 
We further consider that the dominant interaction 
is an $s$-wave one since we operate in the ultracold regime.
As such both intra- ($g_{BB}$, $g_{II}$) and inter-species ($g_{BI}$) interactions are adequately 
described by a contact potential \cite{Olshanii}, see also Appendix~\ref{dimensional_reduction}. 
Furthermore, $\hat{H}_{BB}=(g_{BB}/2) \int dx~\hat{\Psi}^{\dagger}_{B}(x) \hat{\Psi}^{\dagger}_{B}(x) 
\hat{\Psi}_{B} (x)\hat{\Psi}_{B}
(x)$ and $\hat{H}_{aa'}=g_{II}\int dx \hat{\Psi}_a^{\dagger}(x)\hat{\Psi}_{a'}^{\dagger}(x)
\hat{\Psi}_{a'}(x)\hat{\Psi}_a(x)$, with $a,a'\in \left\{ \uparrow, \downarrow \right \}$, 
correspond to the contact intraspecies 
interaction terms of the bosonic bath and the impurities respectively. 
Note that only the spin-$\uparrow$ component of the impurities interacts with the 
BEC while the spin-$\downarrow$ one is noninteracting. 
The relevant interspecies interaction term reads
$\hat{H}_{BI}=g_{BI}\int dx~\hat{\Psi}^{\dagger}_{B}(x) \hat{\Psi}^{\dagger}_{\uparrow}(x) 
\hat{\Psi}_{\uparrow}(x)\hat{\Psi}_{B}(x)$.  
Finally, $\hat{H}_S^{\beta}=\frac{\hbar \Omega_{R0}^{\beta}}{2} \hat{S}_x - \frac{\hbar \Delta^{\beta}}{2} \hat{S}_z$,
with $\Omega_{R0}^{\beta}$ and $\Delta^{\beta}=\nu^{\beta} -\nu_0$ referring to the bare Rabi frequency and the detuning 
of the radiofrequency pulse when the bosonic bath is absent, see Appendix~\ref{rf_implement} for further details.  
Here, $\beta \in \left\{ \rm{pump}, \rm{probe}, \rm{dark} \right \}$. 
Moreover, the total spin operators are given by 
$\hat{{S}_i}=\int dx \sum_{ab} \hat{\Psi}_a (x) 
\text{$\sigma$}_{ab}^i \hat{\Psi}_b (x)$, with $\sigma_{ab}^i$ denoting the Pauli matrix $i\in \{ x,y,z \}$. 

It is worth mentioning at this point that the one-dimensional description adopted holds under the 
conditions $\frac{k_BT}{\hbar \omega} \ll \frac{\hbar^2}{m}[\rho^{(1)}_B (x=0) ]^2 \approx \frac{3^{4/3}}{16} 
( \frac{\alpha_{\perp}^2 N_B^2}{a_{BB} \alpha} )^{2/3} $ and 
$\frac{N_B a_{BB} \alpha_{\perp}}{\alpha^2} \ll 1 $~\cite{Strigari,Pethick}. 
In these expressions, $a_{BB}$ is the three-dimensional $s$-wave scattering length between the particles of the medium, and
$\alpha=\sqrt{\frac{\hbar}{m \omega}}$ and $\alpha_{\perp}=\sqrt{\frac{\hbar}{m \omega_{\perp}}}$ correspond to 
the axial and transversal length scales. 
$\rho^{(1)}_B (x=0)$ is the initial one-body density of the environment at $x=0$, $k_B$ is the Boltzmann constant and 
$T$ refers to the temperature of the bosonic bath. 
To provide a concrete example, assuming $\omega \approx 2\pi \times 100~$Hz and considering a ${}^{87}$Rb gas with 
$g_{BB}=0.5\sqrt{(\hbar^3\omega)/(m)}\approx 3.55 \times 10^{-38}~$Jm our 1D setting can be realized for 
transverse frequencies $\omega_{\perp}\approx 2\pi\times 5.1~$kHz. 
Accordingly, the 1D treatment is valid since $N_B a_{BB} \alpha_{\perp}/\alpha^2 \approx 0.07 \ll 1$ and temperature effects are 
negligible for $k_BT \ll 316\hbar \omega \approx 1.5~\mu K$. 

To access the time-resolved spectral response of bosonic and fermionic impurities immersed 
in the BEC bath the Multilayer Multi-Configuration Time-Dependent Hartree method for atomic 
mixtures is utilized~\cite{MLX,MLB1,MLB2}. 
The latter is a nonperturbative approach that uses a variationally optimized time-dependent 
basis which spans the optimal subspace of the Hilbert space at each time instant and allows for 
tackling all interatomic correlations~\cite{Us_Busch}. 
In particular, the MB  wavefunction is expressed as a truncated Schmidt decomposition using $D$ 
species functions for each component [Eq.~(\ref{eq:wfn}) in Appendix~\ref{sec:method}]. 
Next, each of these species functions is expanded in a basis of $d^B$ and $d^I$ single-particle functions 
for the BEC background and the impurities respectively [Eq.~(\ref{eq:number_states})]. 
These single-particle functions utilize a time-independent primitive basis that is 
a tensor product of basis states regarding the spatial and the spin degrees of freedom [Eq.~(\ref{eq:spfs})]. 
Then, by following a variational principle we arrive at a set of coupled nonlinear integrodifferential equations of 
motion~\cite{MLX,MLB1,MLB2}. 
A detailed description of our MB variational approach and the ingredients of our numerical 
simulations are provided in Appendices~\ref{sec:method} and~\ref{convergence} respectively.

\section{Pump-Probe Spectroscopy Scheme}\label{PPS_details}

\begin{figure*}
\includegraphics[width=1.0\textwidth]{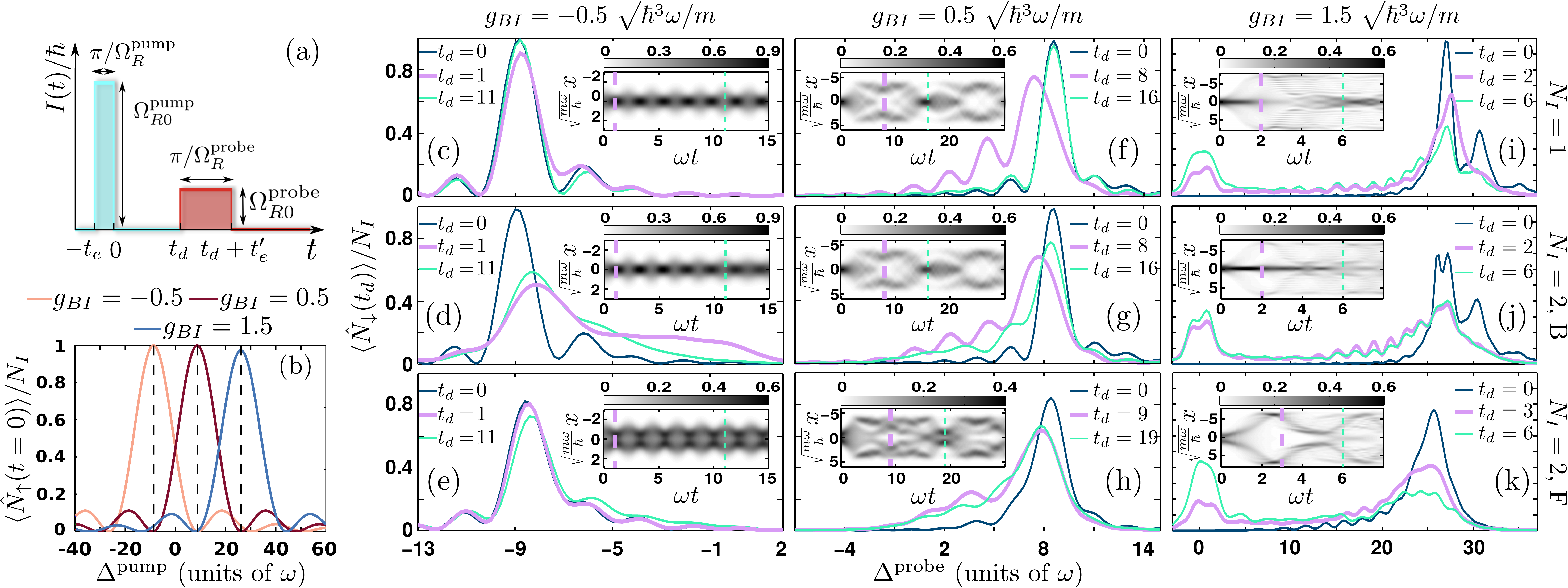}
\caption{(a) Schematic illustration of the PPS pulse sequences used.
(b) Spectral response of the pump pulse $\langle \hat N_{\uparrow}(t=0) \rangle/N_I$ versus its 
detuning $\Delta^{\rm pump}$ for $g_{BB}=0.5\sqrt{\hbar^3 \omega/m}$, $N_{B}=100$, $N_I=1$ 
and varying $g_{BI}$.
Vertical dashed lines indicate the resonant detunings $\Delta_+$. 
(c)-(k) Time-resolved probe spectra at different $g_{BI}$, bosonic (B) or fermionic (F) impurity 
numbers $N_I=1,2$ with $g_{II}=0$, and  
for various short dark times, $t_d$ (see legend). 
In all cases insets illustrate the spatiotemporal evolution of the impurity's one-body density and 
dashed lines mark the instants for which the probe spectrum is provided.} 
\label{Fig:1}
\end{figure*}

We prepare the multicomponent system in its ground state with fixed $g_{BB}$ and $g_{II}=0$. 
The impurities are in their spin-$\downarrow$ state and thus $\langle\hat{H}_{BI}\rangle=0$. 
To trigger the dynamics, an intense,
$\Omega^{\rm pump}_{R0}=10\omega \gg \omega$, rectangular pump pulse 
drives the noninteracting spin-$\downarrow$ impurities to their interacting with the bath spin-$\uparrow$ 
state for $-t_e<t<0$ (where $t_e$ denotes the exposure time) [Fig.~\ref{Fig:1}(a)]. 
The condition $\Omega_{R0}^{{\rm pump}} \gg \omega$ ensures that the duration of the pump pulse 
is much smaller than the time interval in which the polarons form. 
Accordingly, the polaron formation can only occur after the termination of the pump pulse and therefore 
it can be captured by the subsequent probe pulse. 
To ensure the resonance condition of the pump pulse, namely
$\Delta^{\rm pump}=\Delta_+$, and to optimize $t_e=\pi/\Omega^{\rm pump}_R$, the fraction of impurity
atoms that have been successfully transferred to the spin-$\uparrow$ state,
$\langle N_{\uparrow} (t=0) \rangle/N_I$, is monitored for variable $\Delta^{\rm pump}$. 
The resulting pump spectrum features a coherent atomic 
resonance~\cite{Nobeler,Massignan3,Scazza,Us_Fermi,Mistakidis_bose_pol} at 
$\Delta^{\rm pump}=\Delta_{+}$. 
The latter, for $N_I=1$ and $g_{BI}=\pm 0.5, 1.5\sqrt{\hbar^3\omega/m}$, is clearly 
visible in Fig.~\ref{Fig:1}(b). 
Notice also that secondary peaks possessing an intensity of the order of $12\%$
with respect to the dominant ones also emerge due to the rectangular shape of the
pump pulse [see also Appendix~\ref{rf_implement}].  

After the initial pump sequence the remaining population of the
spin-$\downarrow$ state is annihilated by employing an optical blast that projects the impurities to 
the $\ket{\uparrow}$ state (as described in Appendix~\ref{rf_implement}) 
and subsequently the spin-$\uparrow$ atoms are left to evolve for fixed $g_{BI}$ 
and $\Omega^{\rm dark}_{R0}=0$ but variable dark time $t_d$. 
The polaronic states can form within $0\leq t \leq t_d$ while at $t=t_d$ a probe pulse is applied. 
This pulse is characterized by $\Omega^{\rm probe}_{R0}=\omega \ll \Omega^{\rm pump}_{R0}$ 
so as to enhance the spectral resolution of the signal 
obtained by the fraction of impurity atoms transferred to the
spin-$\downarrow$ state, $\langle N_{\downarrow}(t_d) \rangle/N_I$ for variable $\Delta^{\rm probe}$.
For the same reason the duration of the probe pulse is fixed to $t'_e=\pi/\Omega^{\rm probe}_{R}$ where
$\Omega^{\rm probe}_{R}$ is the resonant Rabi frequency of the probe pulse at $\Delta^{\rm
probe}=\Delta_+$, $t_d=0$ and $N_I=1$. 

Concluding within the PPS scheme, polaronic states can be identified in the probe spectrum as well-defined peaks
with amplitude $\langle \hat{N}_{\downarrow}(t_d) \rangle/N_I < 1$. 
For our purposes (accounting for the finite fidelity resulting after the probe pulse) we employ the criterion 
$\langle \hat{N}_{\downarrow}(t_d) \rangle/N_I < 0.96$ in order to identify the polaronic resonances. 
Interestingly, a peak with $\langle \hat{N}_{\downarrow}(t_d) \rangle/N_I \approx 1$ does not correspond to a polaron as it implies
that the accessed MB state is equivalent to a non-interacting state. 
Accordingly, the peaks exactly at $t_d=0$ and $\langle \hat{N}_{\downarrow}(t_d) \rangle/N_I \approx 1$, that will appear later on, 
do not indicate the
formation of polarons. 
Notice that polaronic peaks with $\langle \hat{N}_{\downarrow}(t_d) \rangle/N_I < 1$ 
can occur, for strong impurity-BEC interactions $g_{BI}>g_{BB}$, even for $t_d=0$ demonstrating 
fast energy transfer to the polaronic states 
for $t_d < (\Omega^{\rm probe}_{R0})^{-1}$. 

\subsection{Short-time dynamics of Bose polarons}\label{short_time} 

The short-time ($t\sim \omega^{-1}$) dynamics of few, $N_I=1,2$, fermionic or bosonic impurities with $g_{II}=0$
immersed in a BEC bath of $N_B=100$ atoms, is captured by the probe spectra  
for distinct attractive [Figs.~\ref{Fig:1}(c)-(e)] and repulsive [Figs.~\ref{Fig:1}(f)-(k)]
interspecies interactions $g_{BI}$.
Focusing on the attractive side, a well-defined polaron at $t_d=0$ with central
peak location at $\Delta_{+}=-8.7\omega$ [Fig.~\ref{Fig:1}(c)], 
$\Delta_{+, B}=-8.9\omega$ [Fig.~\ref{Fig:1}(d)], 
and $\Delta_{+, F}=-8.6\omega$ [Fig.~\ref{Fig:1}(e)] is identified 
in the cases of $N_I=1$, and $N_I=2$ bosonic and fermionic impurities respectively.
These polarons show a nonsizeable shift for all the different evolution times $t_d$
as long as $N_I=1$.
However, a clear shift can be inferred for $N_I=2$ [Fig.~\ref{Fig:1}(d)].  
This shift, being of about $10\%$, is a consequence of the energy redistribution between the 
bosonic impurities and the BEC as demonstrated in~\cite{Mistakidis_bose_pol}
and it is further related to the fact that for $g_{BI}<0$ attractive induced-interactions
are significantly enhanced~\cite{Mistakidis_bose_pol}.  
Additionally, due to the pronounced induced-interactions a collisional
broadening \cite{Scully} of the spectral line is clearly observed for $t_d=1,11\omega^{-1}$. 
Indeed, since the two-body state of the impurities evolves rapidly during the probe
sequence the spectral resolution of the measurement decreases giving rise to a
wide background in the PPS spectrum for $-10< \Delta^{\rm probe}/\omega<2$.
The imprint of induced-interactions in the spatiotemporal evolution of the one-body density 
is the dephasing of the breathing oscillations [hardly visible in the inset of Fig.~\ref{Fig:1}(d)] 
within the time interval $10<\omega t_d<15$,    
which is absent for the single impurity [see the inset in Fig.~\ref{Fig:1}(c)].  
In contrast to the above dynamics, the time-resolved evolution of fermionic impurities closely resembles the single 
impurity one with the two fermions undergoing at short-times 
a coherent breathing motion, as is evident in the inset of Fig.~\ref{Fig:1}(e). 
This latter result can be easily understood by the fact that attractive induced-interactions 
between fermionic impurities are known to be suppressed providing in turn a nonsizeable shift 
of the respective atomic peak resonance captured by the probe spectra~\cite{Scazza,Us_Fermi}.

Switching to repulsive interactions, the dynamical evolution of the system
changes dramatically. 
Independently of flavor and concentration the motion of the impurities, as detected by the one-body density 
evolution for $g_{BI}=0.5\sqrt{\hbar^3 \omega/m}$, is apparently qualitatively similar [insets in Figs.~\ref{Fig:1}(f)-\ref{Fig:1}(h)].
From the very early stages of the nonequilibrium dynamics the density filamentizes
with recurrences of an almost central density peak occurring at the collision points, 
i.e. around $t_d\approx 16\omega^{-1}$ and $t_d\approx 19\omega^{-1}$ for the bosonic
and fermionic impurities respectively [Fig.~\ref{Fig:1}(g) and Fig.~\ref{Fig:1}(h)].
However in all three cases a clean quasiparticle peak is monitored in the respective probe spectra
indicating the existence of well-defined polarons for $t\sim \omega^{-1}$.
The dominant peak location appears to be shifted for $t_d\neq 0$ when compared to $\Delta_{+}$, 
while an overall broadening of the spectrum is also inferred.
The observed shift is independent of the number of 
impurities [compare Figs.~\ref{Fig:1}(f) and \ref{Fig:1}(g)] 
but depends strongly on the impurity's nature with the shift measured to be of about $10\%$ for 
bosonic but dropping down to almost $5\%$ for fermionic impurities [Figs.~\ref{Fig:1}(g), \ref{Fig:1}(h)].
These findings suggest that attractive induced-interactions cannot be directly 
unveiled by the observed shift. 
A result that complements earlier predictions indicating that attractive 
induced-interactions are suppressed in the repulsive regime~\cite{us,us1}. 
Indeed, for $g_{BI}>0$ the density of the BEC is less distorted compared to $g_{BI}<0$ \cite{us,Sdiss}. 
Accordingly, the impurities are less attracted to these distortions and consequently to each other. 
In turn by inspecting the oscillatory tail of the probe spectra
for $g_{BI}>0$ interference phenomena associated with the filamentation process can be identified.
Indeed, already from the single impurity [Fig.~\ref{Fig:1}(f)]
the amplitude, $A(\Delta^{\rm{probe}})$, of the secondary peak appearing in the spectra e.g. at $t_d=8\omega^{-1}$, 
$A(\Delta^{{\rm probe}}\approx4.6\omega)=0.352$ is 
comparable with the dominant one $A (\Delta^{{\rm probe}}\approx7.5\omega)= 0.755$. 
Notice that the intensity ratio of the secondary to the dominant peak is larger that $12\%$ and 
thus cannot be attributed to the rectangular shape of the probe pulse. 
The latter, directly reflects the coherence between 
the filaments formed in the one-body density (see discussion below).
However, as the number of impurities increases significant losses of coherence take place.  
Indeed, the secondary peak at $t_d=8\omega^{-1}$ [$t_d=9\omega^{-1}$] has $A(\Delta^{\rm probe}\approx 4.6\omega)=0.266$ 
[$A(\Delta^{\rm probe}\approx3.6\omega)=0.245$] 
for $N_I=2$ bosonic [fermionic] impurities while is drastically reduced at later $t_d$ [Figs.~\ref{Fig:1}(g), \ref{Fig:1}(h)]. 
These losses of coherence are an indirect manifestation of the presence of weak attractive induced-interactions which we cannot 
probe via the shift of the spectral peaks. 

Our PPS data demonstrate that well-defined quasiparticles cease to exist for $g_{BI}\gtrsim 1.5\sqrt{\hbar^3 \omega/m}$ 
signaling their TOC [Figs.~\ref{Fig:1}(i)-(k)]~\cite{Mistakidis_bose_pol}. 
Evidently, at $t_d=0$ a predominant peak  
centered at $\Delta_+=26.7\omega$
can be discerned in the single impurity probe spectrum [Fig.~\ref{Fig:1}(i)], 
giving its place to a double humped structure with averaged location at $\Delta_{+,B}=26.8\omega$ 
for the two bosonic impurities [Fig.~\ref{Fig:1}(j)], 
and to a slightly shifted but significantly broadened 
peak at $\Delta_{+,F}=25.6\omega$ for the fermionic ones [Fig.~\ref{Fig:1}(k)]. 
The latter broadening is attributed to the fermion statistics. 
Indeed fermionic impurities occupy higher momenta and as such couple stronger to the BEC excitations. 
However, at $t_d=2\omega$ deformation of the central peak is 
present and a highly oscillatory tail is seen in all cases.
To appreciate the aforementioned degree of coherence already indicated by the 
probe spectra we next invoke the spatial first order coherence function
\begin{equation}
g^{(1)}_{\sigma}(x,x';t)=\frac{\rho^{(1)}_{\sigma}(x,x';t)}{\sqrt{\rho^{(1)}_{\sigma}(x;t) 
\rho^{(1)}_{\sigma}(x';t)}}.\label{coherence_first} 
\end{equation}
Here, $\rho^{(1)}_{\sigma}(x,x';t)=\braket{\Psi(t)|\hat{\Psi}^{\dagger}_{\sigma}(x)\hat{\Psi}_{\sigma}(x')|\Psi(t)}$ is the 
$\sigma$-species ($\sigma=B,\uparrow,\downarrow$) one-body reduced density matrix, $\ket{\Psi(t)}$ is the MB 
wavefunction and $\rho^{(1)}_{\sigma}(x;t)$ is the one-body density, see also Appendix~\ref{sec:method}.  
Importantly, $|g^{(1)}_{\sigma}(x,x';t)| \in [0, 1]$ 
indicates the spatially resolved deviation of a MB wavefunction from a corresponding product state. 
Specifically, if $|g^{(1)}_{\sigma}(x,x';t)|=1$ the system is termed 
fully coherent otherwise coherence losses occur 
signifying the build-up of correlations~\cite{us,us_phase_sep}. 
Indeed, the instantaneous $|g^{(1)}_{\uparrow}(x,x';t_d=2\omega)|$ for $g_{BI}=1.5\sqrt{\hbar^3 \omega/m}$ clearly 
dictates that the quasiparticle remains adequately coherent 
\begin{figure}[ht]
\includegraphics[width=1.0\columnwidth]{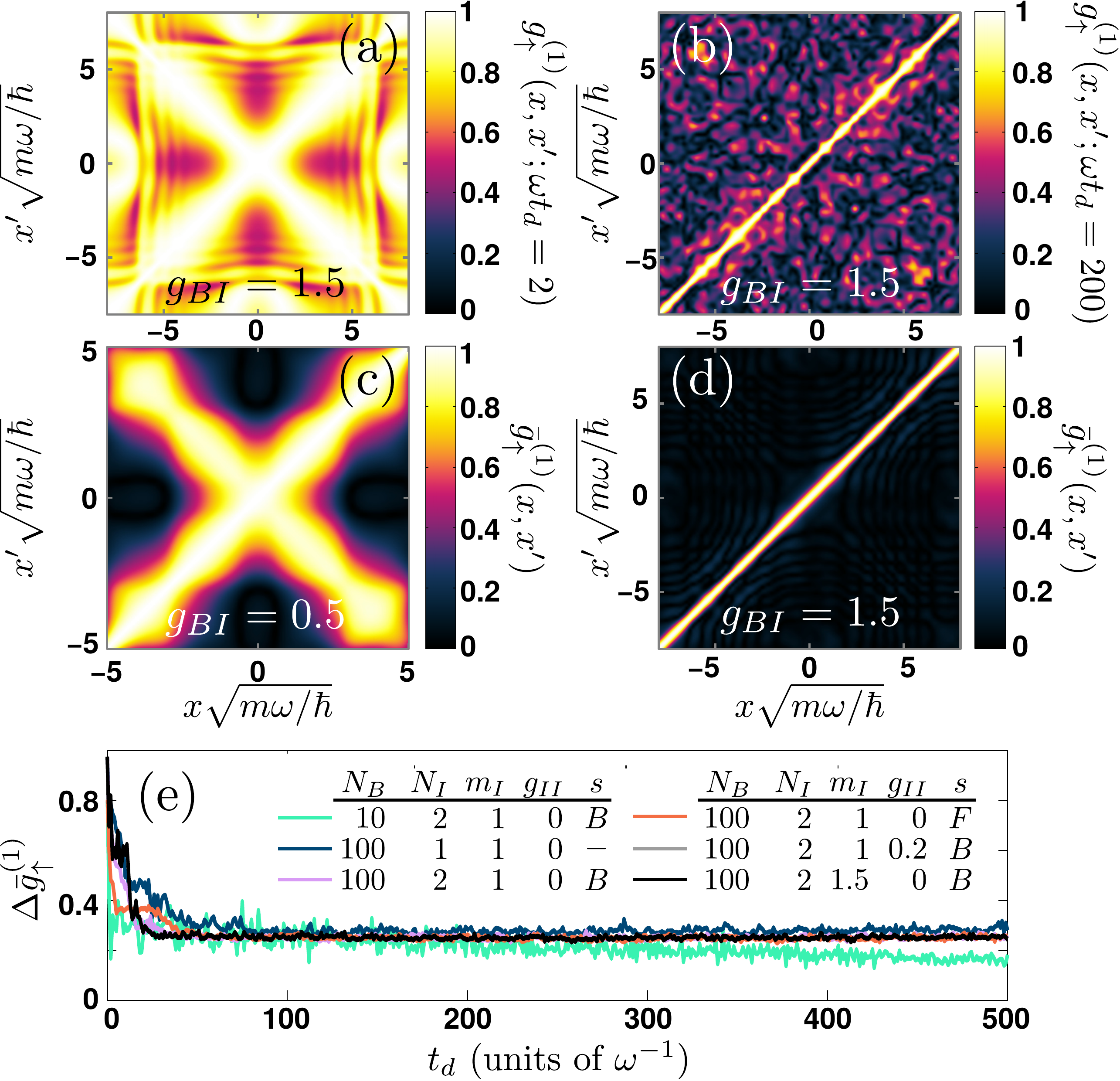}
\caption{(a), (b) One-body coherence $g^{(1)}_{\uparrow}(x,x';t_d)$
at different times $t_d$ (see legend).
(c), (d) Time-averaged one-body coherence $\bar{g}^{(1)}_{\uparrow}(x,x')$ at distinct $g_{BI}$.
(e) Temporal evolution of the variance $\Delta\bar{g}^{(1)}_{\uparrow}$ for different settings (see legend).}
\label{Fig:2}
\end{figure}
since e.g. $|g^{(1)}_{\uparrow}(x=-5\sqrt{\hbar/m\omega},x'=5\sqrt{\hbar/m\omega};t_d=2\omega)|\approx 0.96$ 
[Fig.~\ref{Fig:2}(a)]. 
Finally, notice that for $t_d>6\omega$ any quasiparticle notion is lost as detected by the probe spectra.
This outcome, being consistent with recent works~\cite{Mistakidis_bose_pol,Us_Fermi}, 
is also supported by the diffusive behaviour of the corresponding one-body density evolution 
[insets in Figs.~\ref{Fig:1}(i)-\ref{Fig:1}(k)]. 
In contrast, a peak corresponding to free particles $\Delta^{\rm probe}=0$
emanates in the PPS spectrum, referring to a phase separation between the impurity and the BEC. 
It is also worth mentioning that by employing a corresponding Ramsey scheme [see the discussion in Appendix~\ref{Ramsey}] 
it is not possible to conclude the emergence of the TOC within the same time interval since the structure factor is still finite and 
drops close to zero for substantially longer evolution times.

\subsection{Long-time Bose polaron dynamics}\label{long_time}

\begin{figure}[ht]
\includegraphics[width=1.0\columnwidth]{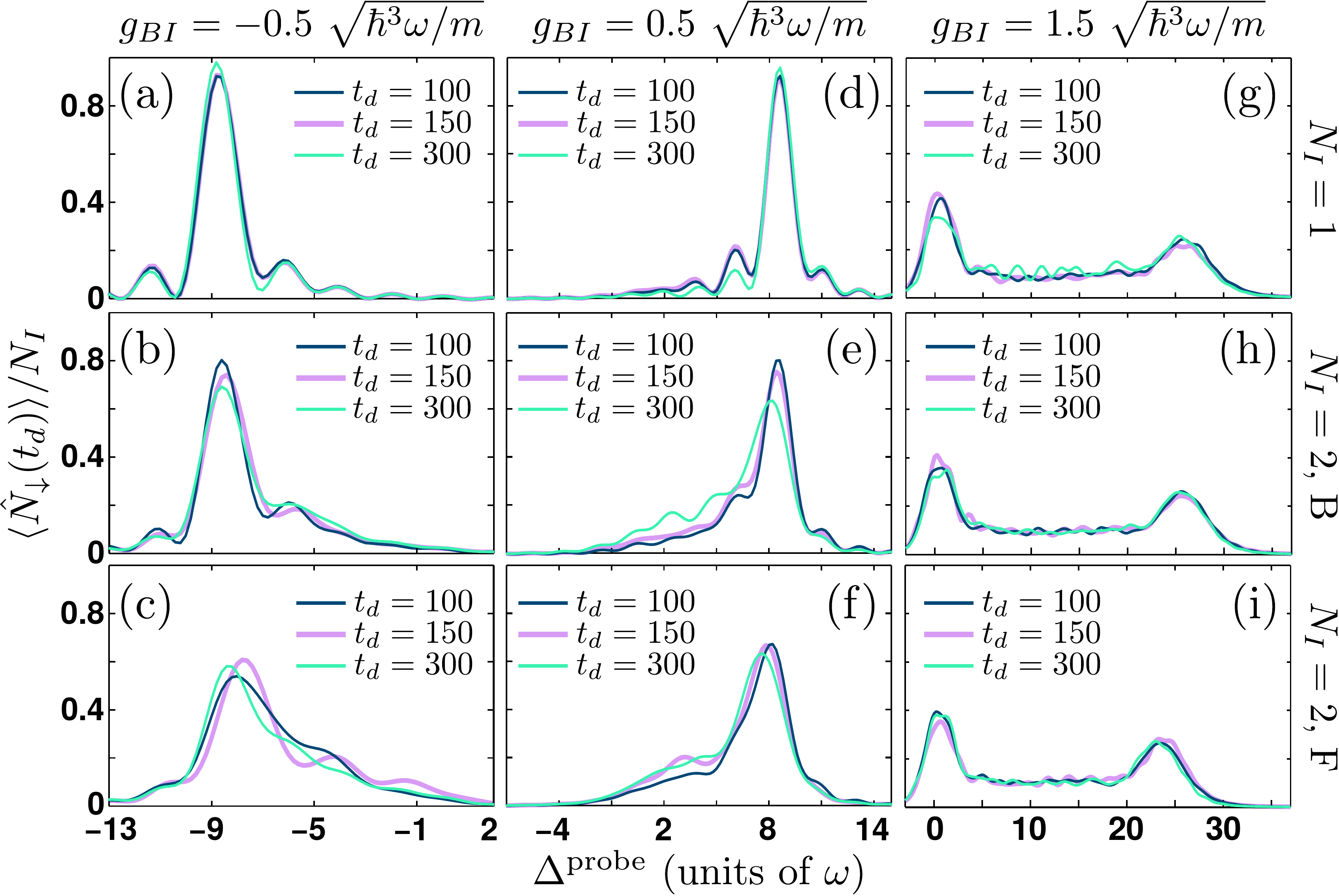}
\caption{(a)-(i) Probe spectra for different $g_{BI}$, $N_I$, 
and impurity flavors at various dark times $t_d$ 
deep in the evolution (see legend). 
The remaining system parameters are the same as in Fig.~\ref{Fig:1}.} 
\label{Fig:3}
\end{figure}

Next let us study the evolution of the system at longer times $100<\omega t_d<300$. 
Note that for a typical axial confinement $\omega \approx 2\pi \times 100~$Hz, the interval 
$100<t_d<300$ corresponds to $0.16<t_d<0.48$~seconds. 
As time evolves one expects that significant losses of coherence signaling the
build-up of correlations will take place in the MB evolution of the system [Fig.~\ref{Fig:2}(b)].
A powerful asset of exposing the latter is the temporal average 
\begin{equation}
\bar{g}^{(1)}_{\uparrow} (x,x') = \lim _{T\to \infty}\frac{1}{T}\int_{0}^{T} dt g^{(1)}_{\uparrow} (x,x';t), \label{coh_avg}
\end{equation} 
which depends solely on the eigenstate properties of the interacting system \cite{Mistakidis_expansion}. 
This allows us to infer the relaxation tendency of the impurities in the framework of the 
ETH~\cite{Rigol1,Rigol2}, see also the discussion in Sec.~\ref{relaxation_ETH}.
Evidently, $\bar{g}^{(1)}_{\uparrow} (x,x')$ reveals that  
for $g_{BI}=0.5\sqrt{\hbar^3 \omega/m}$ the impurity is largely coherent [Fig.~\ref{Fig:2}(c)]
whilst at $g_{BI}=1.5\sqrt{\hbar^3 \omega/m}$ any coherence property is lost [Fig.~\ref{Fig:2}(d)].
This outcome is further supported by the time-resolved probe spectra illustrated for longer 
times in Figs.~\ref{Fig:3}(a)-\ref{Fig:3}(i).
A strong suppression of the interaction shift with respect to $\Delta_+$ is found to persist
at long times which together with the weakly fluctuating amplitude $A(\Delta^{\rm probe}\approx \Delta_+)$ observed 
in the course of time verifies the longevity of coherent single and two polarons irrespectively of their flavor and 
for both attractive and moderate repulsive $g_{BI}=\pm 0.5\sqrt{\hbar^3 \omega/m}$ [Figs.~\ref{Fig:3}(a)-\ref{Fig:3}(f)]. 
Alterations come into play for $g_{BI}>g_{BB}$ where as per our discussion above 
losses of coherence, as captured by ${g}^{(1)}_{\uparrow} (x,x';t_d)$, become significant and the 
polaron picture breaks down. 
Here, a two-humped distribution appears in our probe spectra independently of the number of the impurities 
and their nature. 
The most pronounced feature of Figs.~\ref{Fig:3}(g)-\ref{Fig:3}(i) is the peak located at $\Delta^{\rm probe}=0$. 
The latter findings suggest that a relaxed state is reached 
characterized by incoherent impurities being unpredicted so far. 

Indeed, by fitting $\bar{g}^{(1)}_{\uparrow}(x,x')$ to the corresponding prediction 
of the $N_I$-particle Gibbs ensemble we obtain large effective
temperatures.
These refer to $k_B T_{\rm eff}=8.45\hbar \omega$ for $N_I=1$ 
and $k_B T_{\rm eff}=8.58\hbar \omega$ ($k_B T_{\rm eff}=5.89\hbar \omega$) 
in the case of two bosons (fermions) showcasing their tendency 
to approach an incoherent thermalized state, see also our detailed discussion in Secs.~\ref{relaxation_ETH} and 
\ref{sec:interaction_energy}. 
Notice that the initial state of fermions involves higher momenta than bosons, while the critical velocity of 
the BEC is the same \cite{velocity1,velocity2,velocity3}. 
Therefore, fermions couple stronger to the BEC excitations losing a larger portion of their energy 
implying a smaller $T_{\rm eff}$. 
Further evidences supporting the observed thermalization~\cite{Lausch1,Lausch2}
are provided by the temporal evolution of the variance 
\begin{equation}
\begin{split}
\Delta \bar{g}^{(1)}_{\uparrow} =\frac{1}{T 
S} \int_S dx dx' \int dt \left[g^{(1)}_I (x,x';t) - \bar{g}^{(1)}_I (x,x') \right]^2, \label{coh_variance} 
\end{split}
\end{equation}
with $S$ denoting the relevant spatial region in which the impurities reside 
and $\Delta \bar{g}^{(1)}_{\uparrow}\in (0,1)$. 
Remarkably a tendency towards thermalization is seen [Fig.~\ref{Fig:2}(e)], 
with $\Delta \bar{g}^{(1)}_{\uparrow}$ saturating at long times irrespectively 
of the size of the BEC cloud and whether one or two, noninteracting or interacting, impurities are present and what their nature is.   

\section{Impurity-medium interaction energy} \label{sec:interaction_energy}

To further support the thermalization tendency of the multicomponent system for strong impurity-BEC 
interactions at longer times of the nonequilibrium dynamics, we next inspect the behavior 
of the interspecies interaction energy. 
The latter quantity is defined as $\braket{\hat{H}_{BI}(t)}\equiv \braket{\Psi(t)|\hat{H}_{BI}|\Psi(t)}$, where the 
operator of the interspecies interactions is 
$\hat{H}_{BI}=g_{BI}\int dx~\hat{\Psi}_B^{\dagger}(x) \hat{\Psi}_{\uparrow}^{\dagger}(x) \hat{\Psi}_{\uparrow}(x)\hat{\Psi}_B(x)$. 
Also, $\hat{\Psi}_{\sigma} (x)$ and $\hat{\Psi}_{\sigma}^{\dagger} (x)$ denote the $\sigma$-species 
field operator that annihilates and creates respectively a $\sigma$-species particle 
at position $x$. 
\begin{figure}[ht]
\includegraphics[width=0.95\columnwidth]{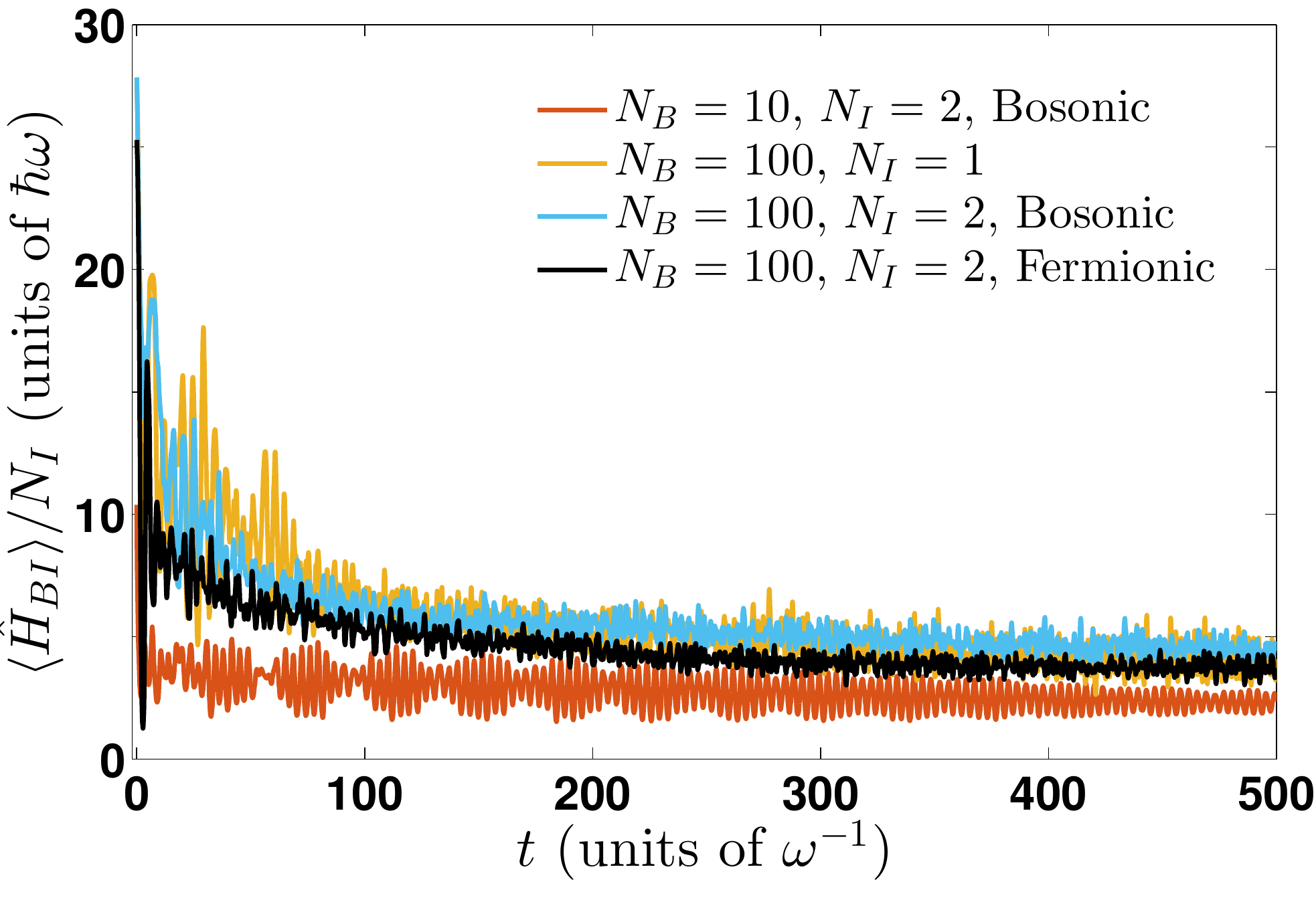}
\caption{Evolution of the impurity-BEC interaction energy per impurity particle applying a pump
pulse to drive the impurities to the spin-$\uparrow$ state with
$g_{BI}=1.5\sqrt{\hbar^3 \omega/m}$ for a single ($N_I=1$) and two ($N_I=2$) bosonic or fermionic impurities and also for 
a few--body bath consisting of $N_B=10$ particles (see legend). 
In all cases $g_{BB}=0.5\sqrt{\hbar^3 \omega/m}$ and $g_{II}=0$.} 
\label{Fig:interspecies_energy}
\end{figure}

The time-evolution of $\braket{\hat{H}_{BI}(t)}/N_I$ is 
illustrated in Fig.~\ref{Fig:interspecies_energy} upon considering 
a pumping that drives the atoms to the spin-$\uparrow$ state with $g_{BI}=1.5\sqrt{\hbar^3 \omega/m}$.
Specifically, we consider different settings consisting of $N_I=1$, $N_I=2$ spin-polarized bosons or fermions as well as a 
few-body system containing $N_B=10$ bosons and $N_I=2$ 
non-interacting impurities. 
In all cases we observe that $\braket{\hat{H}_{BI}(t)}/N_I$ decreases up to $t=100\omega^{-1}$ while for later 
times, and in particular for $t>200\omega^{-1}$, it shows a saturation trend to a certain value depending on 
both $N_I$ and $N_B$. 
Recall that this saturation effect allowed for the derivation of Eq.~(\ref{eq:fer_distr}) within the ETH scheme. 
Notice also that the saturation value of $\braket{\hat{H}_{BI}(t)}/N_I$ is smaller for the few-body 
system ($N_B=10$, $N_I=2$) while for the $N_B=100$ setups $\braket{\hat{H}_{BI}(t)}/N_I$ acquires 
its smaller value for two fermionic impurities and takes almost the same value for $N_I=1$ and 
$N_I=2$ non-interacting bosonic impurities. 
Finally, let us note that the overall 
decreasing behavior of $\braket{\hat{H}_{BI}(t)}/N_I$
suggests a transfer of energy from the impurities to the bosonic gas 
as it has been also demonstrated in Refs.~\cite{Us_Busch,Sdiss}. 
This energy transfer process, identified by the decreasing rate of $\braket{\hat{H}_{BI}(t)}/N_I$, 
seems to be enhanced for $N_B=10$ whilst for the $N_B=100$ case it is more pronounced for the 
fermionic impurities. 
We remark that a saturation trend at long time-scales being in turn suggestive of the thermalization tendency of the 
system occurs also for other observables. 
These include, for instance, the dynamical structure factor [Appendix~\ref{Ramsey}] as well as 
entropic measures~\cite{Bera,Roy} such as the von-Neumann entropy~\cite{Sdiss,us_phase_sep,phase_sep_ferm} 
quantifying the degree of entanglement (results not shown for brevity).

\section{Characterization of the relaxation dynamics}\label{relaxation_ETH}

We next explicate our method for characterizing the relaxation dynamics
occurring in our setup during the hold time $t_d$. To achieve this we employ the ETH~\cite{Rigol1,Rigol2}. 
Within this framework it is assumed that after a quench, the finite subsystems of a larger 
extended system relax to a steady state reminiscent of thermal equilibrium. 
Here by fitting the time-averaged one-body density of the impurities to a
thermal equilibrium one, we show that this thermalization process explains 
the relaxed state of the impurities emanating for long times 
after the orthogonality catastrophe of the polarons.

The relaxation of an isolated (closed) system is understood in terms 
of the principle of local equivalence~\cite{Essler}.
Within this framework as the thermodynamical limit is approached, i.e. the system size
tends to infinity, the reduced density matrices of the involved 
few-particle subsystems at long times can be calculated in terms of the density matrix of a 
(generalized) Gibbs ensemble at thermal equilibrium. 
Indeed, if the only conserved quantity of the Hamiltonian is the total energy then it can
be shown that these reduced density matrices can be calculated in terms of the 
equilibrium density matrix within the Gibbs ensemble
\begin{equation}
    \hat{\rho}_{\rm Gibbs}=\frac{1}{Z} e^{-\frac{\hat{H}}{k_B T_{\rm eff}}}.
    \label{eq:Gibbs}
\end{equation}
In Eq.~(\ref{eq:Gibbs}) $Z$ is the partition function stemming 
from the normalization of the density matrix, 
i.e. ${\rm Tr}\left[\hat{\rho}_{\rm Gibbs}\right]=1$. 
$\hat{H}$ refers to the MB Hamiltonian and $T_{\rm eff}$, $k_B$ 
correspond to the effective temperature and the Boltzmann constant respectively. 
Of course, our setup exhibits also other conserved quantities than the total energy. 
Below we resort to the approximation of no further symmetries 
as it is the only case that explicit results showing the relaxation dynamics of the 
system are available within ETH \cite{Essler}.  
As we shall show later on, the aforementioned choice leads to an excellent agreement between our 
numerical findings and the relevant estimates provided by applying Eq.~(\ref{eq:Gibbs}). 
Within this approximation the effective temperature, $T_{\rm eff}$, 
is fixed by the conserved value of the energy per particle in the thermodynamic limit (TL)
\begin{equation}
    \lim_{\rm BEC \to TL} \tfrac{{\rm Tr}\left[ |\Psi(0)\rangle \langle \Psi(0) | \hat{H}
    \right]}{N_B} =\lim_{\rm BEC \to TL} \tfrac{{\rm Tr}\left[ \hat{\rho}_{\rm Gibbs} \hat{H}
    \right]}{N_B}.
    \label{eq:therm}
\end{equation}
Here, ${\rm BEC \to TL}$ is defined as the limit where 
$N_B \to\infty$, $g_{BB} \to 0$, $N_B g_{BB}={\rm constant}$ and $g_{BI}/g_{BB}={\rm constant}$.  
Notice, however, that Eq.~(\ref{eq:Gibbs}) and Eq.~(\ref{eq:therm}), are impractical for 
calculations since the eigenvalues and eigenstates of the full interacting 
Hamiltonian, $\hat{H}$, are required for the evaluation of the Gibbs ensemble 
of the $\left(N_B+N_I\right)$ MB ensemble which are difficult if not impossible to obtain. 
For this reason, we simplify the above-mentioned set of equations so as to obtain
explicit results which can be subsequently compared with those obtained by the time-evolution of 
the $\left(N_B+N_I\right)$ MB system within the ML-MCTDHX approach.

Since we intend to employ the thermodynamic limit where the MF Gross-Pitaevskii
treatment of the BEC is exact in the weak interaction limit~\cite{Pethick}, it is reasonable to assume that 
the corresponding density operator of the Gibbs ensemble acquires the product form 
$\hat{\rho}_{\rm Gibbs}=\hat{\rho}^{(N_B)}_{B;{\rm Gibbs}} \otimes \hat{\rho}^{(N_I)}_{\uparrow;
{\rm Gibbs}}$. 
Recall that during the dark time all of the impurities are
in their spin-$\uparrow$ state. 
In this case, the form of $\hat{\rho}_{\uparrow;\rm Gibbs}$ is
similar to Eq.~(\ref{eq:Gibbs}), namely
\begin{equation}
 \hat{\rho}^{(N_I)}_{\uparrow;{\rm Gibbs}}=\frac{1}{Z_{\uparrow}} e^{-\frac{\hat{H}^{\rm eff}_{\uparrow}}{k_B T_{\rm eff}}},
    \label{eq:Gibbs_impurity}
\end{equation}
where $H_I^{\rm eff}$ is an effective Hamiltonian that acts only on the impurity. 
Equation~(\ref{eq:Gibbs_impurity}) greatly simplifies the description of our system, 
since the density matrix of the impurity depends only on the eigenvectors 
and eigenvalues of a $N_I$-body effective Hamiltonian. 
To proceed further we specify the form of $\hat{H}^{\rm eff}_{\uparrow}$.
Within a zeroth-order approximation we assume that the BEC acts solely as a potential 
barrier for the impurities and as consequence their effective Hamiltonian reads
\begin{equation}
    \hat{H}^{\rm eff}_{\uparrow}(t)=-\small\sum_{i=1}^{N_I} \tfrac{\hbar^2}{2 m_I} \tfrac{d^2}
    {dx_i^2} + \tfrac{1}{2} m_I
    \omega^2 x_i^2 +
    g_{BI} \rho^{(1)}_B(x_i;t).
    \label{eq:effective_Hamiltonian}
\end{equation}
Notice that this approximation for the effective potential is a simplification of the 
impurity problem. First it neglects, among others, the renormalization of the 
impurity's mass, $m_I \to m^{\rm eff}_I$, due to the presence of the BEC \cite{Us_Busch}. 
Second, the presence of induced-interactions between the impurities 
cannot be captured~\cite{us}. 

The time-dependence of the Hamiltonian of Eq.~(\ref{eq:effective_Hamiltonian}) 
implies a non-stationary state for the impurities. 
However, it is well-known that following an interaction quench the density of the BEC is only 
slightly perturbed by the motion of the impurities \cite{Us_Busch,Sdiss,us}. 
The latter justifies the substitution of the effective Hamiltonian by its 
time-averaged value $\hat{\bar{H}}^{\rm eff}_{\uparrow}= \lim_{T \to\infty} \frac{1}{T} \int dt~
\hat{H}^{\rm eff}_{\uparrow}(t)$, since $\rho^{(1)}_B(x_i;t) \approx
\lim_{T \to \infty} \frac{1}{T} \int_0^T dt~\rho^{(1)}_B(x_i;t)$ for the density of the bath. 
By incorporating the above-mentioned approximations we obtain explicit forms for the one-body 
density of the impurity within the Gibbs ensemble, namely
\begin{equation}
    \rho_{I;{\rm Gibbs}}^{(1)}(x,x';T_{\rm eff})=\sum_{i=1}^{\infty} n_i(T_{\rm eff}) \phi_i(x)
    \phi^{*}_i(x').
    \label{eq:red_den_mat}
\end{equation}
Here, $n_i(T_{\rm eff})$ is the distribution function of the $N_I$ particles and $\phi_i$ refers to 
the eigenstates of $\hat{\bar H}^{\rm eff}_{\uparrow}$ respectively. 
Due to the small number of impurities considered herein ($N_I=1,2$) 
both the fermionic and the bosonic impurities do not follow the appropriate, 
for $N_I \to \infty$, Fermi-Dirac or Bose-Einstein distributions. 
Instead, it can be shown that the relevant distribution in the 
case of a single-particle or two bosons is the Boltzmann 
distribution 
\begin{equation}
n_i(T_{\rm eff})=Z(1)^{-1}\exp(-\frac{\epsilon_i}{k_B T_{\rm eff}}),
\end{equation}
with $\epsilon_i$ being the eigenvalues of $\hat{\bar H}^{\rm eff}_{\uparrow}$ and 
$Z(1)=\sum_{i=1}^{\infty}\exp(-\frac{\epsilon_i}{k_B T_{\rm eff}})$. 
For two-fermions the corresponding distribution reads
\begin{equation}
     n_i(T_{\rm eff})=\left[ \tfrac{Z(2)-e^{\frac{-\epsilon_i}{k_B T_{\rm eff}}}(Z(1)-
     e^{\frac{-\epsilon_i}{k_B T_{\rm eff}}})}{Z(1)-e^{\frac{-\epsilon_i}{k_B T_{\rm eff}}}}
 e^{\tfrac{\epsilon_i}{k_B T_{\rm eff}}}+1 \right]^{-1},
    \label{eq:fer_distr}
\end{equation}
where $Z(2)=\sum_{i=1}^{\infty} e^{-\frac{\epsilon_i}{k_B T_{\rm eff}}}(Z(1)-e^{-\frac{\epsilon_i}
{k_B T_{\rm eff}}})$. 
Note also that the one-body density of the impurity, Eq.~(\ref{eq:red_den_mat}), depends
only on a single parameter namely the effective temperature, $T_{\rm eff}$.
\begin{figure}[h]
    \centering
    \includegraphics[width=0.9\columnwidth]{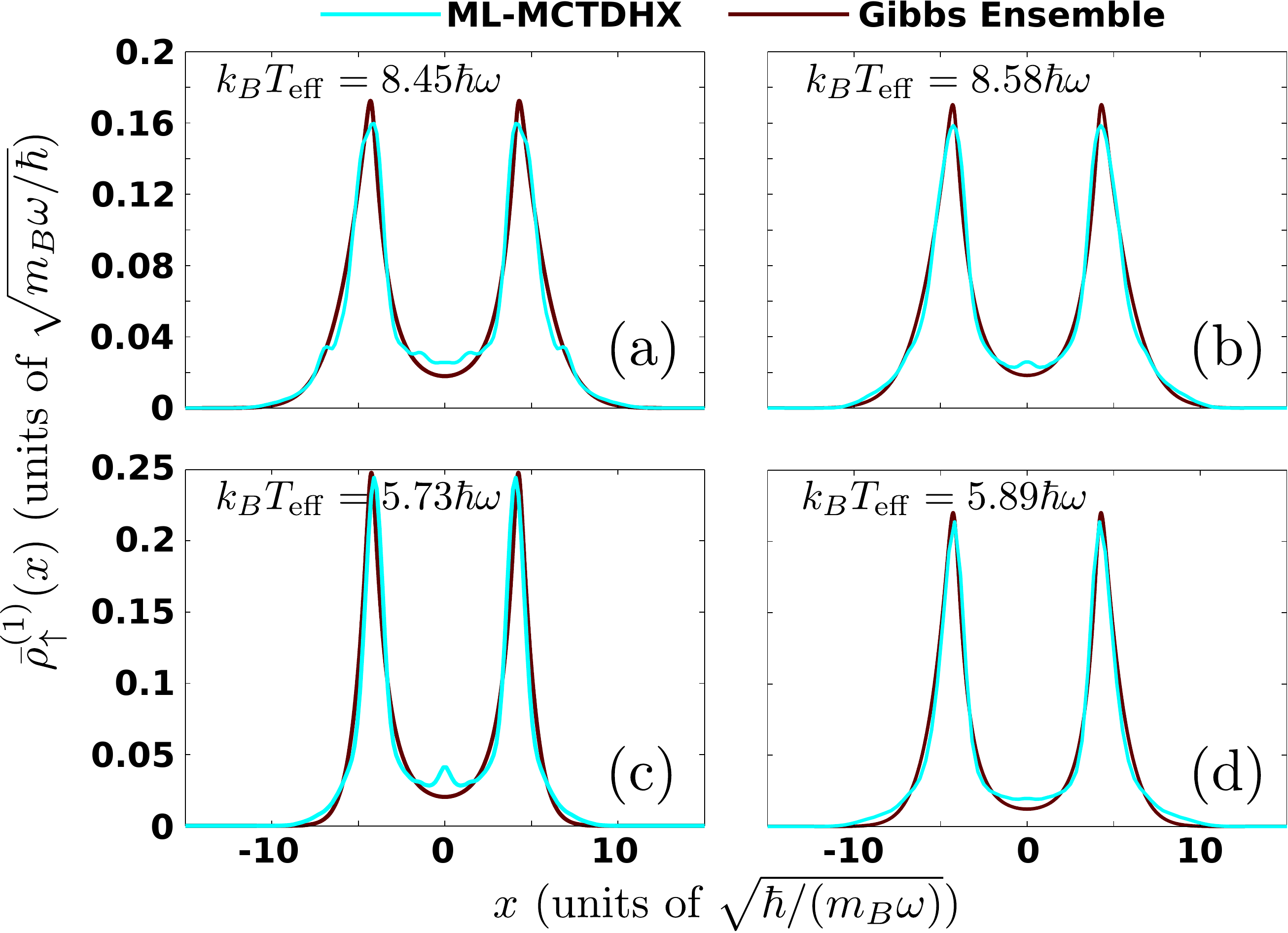}
    \caption{The time-averaged one-body density,
    $\bar{\rho}^{(1)}_{\uparrow}(x)$ within the MB approach (light blue lines)
    compared to the best fit of the ETH model,
    $\rho^{(1)}_{\uparrow;Gibbs}(x;t_{\rm eff})$ (dark red lines). 
    Our results refer to (a) $N_I=1$, (b) $N_I=2$  bosonic and (c) $N_I=2$ fermionic
    mass-balanced, $m_I=m_B$ impurities. 
    (d) Corresponds to the case of $N_I=2$ heavy, $m_I=(133/78) m_B$, bosonic impurities.}
    \label{fig:temperatures}
\end{figure}

As shown earlier the one-body density matrix of the impurities,
$\rho^{(1)}_{\uparrow}(x,x';t)$, saturates to its time-averaged value, i.e.  
$\rho^{(1)}_{\uparrow}(x,x';t \to \infty) \approx \bar{\rho}^{(1)}_I(x,x')=
\lim_{T \to \infty}\frac{1}{T} \int_0^T dt~ \rho_{\uparrow}^{(1)}(x,x';t)$,
for long hold times, as it is evident 
in the relaxation dynamics of $\Delta \bar{g}^{(1)}$ [see also Fig.~\ref{Fig:2}(e)]. 
In order to facilitate the comparison of our results to the ETH prediction [Eq.~(\ref{eq:red_den_mat})] we 
fit the averaged one-body density matrix,
$\bar{\rho}_{\uparrow}^{(1)}(x,x')$, obtained within ML-MCTDHX to the corresponding Gibbs ensemble,
$\rho_{\uparrow;{\rm Gibbs}}^{(1)}(x,x';T_{\rm eff})$, and extract the value of $T_{\rm eff}$. 
Our results for the best fitted parameters are shown in Fig.~\ref{fig:temperatures}. 
We remark that the fitting is performed on the level of 
$\bar{\rho}^{(1)}_{\uparrow}(x,x')$, while only the diagonal $\bar{\rho}^{(1)}_{\uparrow}(x) \equiv
\bar{\rho}^{(1)}_{\uparrow}(x,x)$ is presented in Fig.~\ref{fig:temperatures} in order to enhance 
the visibility of the obtained results.
By comparing the time averaged one-body density and the fitted Gibbs ensemble prediction a
very good agreement is observed, for both one [Fig.~\ref{fig:temperatures}(a)] and two impurities
of either bosonic [Fig.~\ref{fig:temperatures}(b)] 
or fermionic [Fig.~\ref{fig:temperatures}(c)] nature. 
This result holds equally also in the case of mass-imbalanced mixtures composed for instance of 
heavy bosonic impurities, $m_I=133/78 m_B$ [see e.g. Fig.~\ref{fig:temperatures}(d)]. 
The above findings indicate that despite the employed approximations the ETH scheme 
is able to capture the main features exhibited by the relaxed state of the impurities 
within our correlated MB system.

Regarding the effective temperature we find rather large values of $T_{\rm eff}$ 
for one and two bosonic impurities that is of 
the order of $T_{\rm eff}\approx8 \hbar \omega/k_B$ harmonic units. 
Whilst for two fermions and two heavier bosonic impurities 
the temperature is slightly smaller possessing values 
of the order of $T_{\rm eff}\approx 6 \hbar \omega/k_B$. 
These large values of the effective temperature are indicative of the incoherent 
character of the impurities after the probe pulse [see also Fig.~\ref{Fig:2}(d)]. 
To advance further the correspondence between the ETH model and the correlated MB
results we estimate the effective temperature of the relaxed state by 
expressing Eq.~(\ref{eq:therm}) only in terms of the impurity's degrees of freedom. 
Notice that the energy of the impurity is not conserved during the MB evolution 
of our system due to the presence of energy exchange processes 
between the impurity and the bath. 
However, as evidenced in Fig.~\ref{Fig:interspecies_energy}, 
the energy of the impurities saturates for large times 
(see also Section~\ref{sec:interaction_energy}). 
Indeed, by taking advantage of this observation we can cast Eq.~(\ref{eq:therm}) 
in the form
\begin{equation}
    \begin{split}
        \bar{E}_{\uparrow}=\lim_{T \to \infty} \frac{1}{T}\int_0^T dt~\langle \Psi(t) | \hat{H} - \hat{H}^0_B -\hat{H}_{BB} | \Psi(t)
    \rangle\\
    ={\rm Tr}\left[ \hat{\rho}^{(N_I)}_{\uparrow;{\rm Gibbs}} \hat{\bar{H}}_{\uparrow}^{\rm eff} \right],
    \end{split}
    \label{eq:approx_temp}
\end{equation}
where $\bar{E}_{\uparrow}$ is the time-averaged impurity energy. 
In the case of a single impurity, Eq.~(\ref{eq:approx_temp}) 
gives an estimation for the effective temperature of $T_{\rm eff}=8.56 \hbar \omega/k_B$
which is in good agreement with the effective temperature obtained 
by fitting $T_{\rm eff}=8.45 \hbar \omega/k_B$. 
Note also here that the effective Hamiltonian of Eq.~(\ref{eq:effective_Hamiltonian}) 
is known to overestimate the zero-point energy of the impurity
since it neglects its dressing by the excitations of the BEC~\cite{Us_Busch}. 
This in turn explains the higher $T_{\rm eff}$ obtained via Eq.~(\ref{eq:approx_temp}). 
In contrast, in the case of two impurities the related 
estimates for $T_{\rm eff}$ are much higher than the ones obtained
by the fitting of $\bar{\rho}_{\uparrow}^{(1)}(x,x')$.
Specifically, Eq.~(\ref{eq:approx_temp}) yields 
$T_{\rm eff}=10.28 \hbar \omega/k_B$ and $T_{\rm eff}=6.89 \hbar \omega/k_B$ for the two
bosonic and the two fermionic impurities respectively. \
The observed discrepancy is attributed to the presence of induced interactions 
between the impurities that are more prevalent in the case of
bosonic impurities than fermionic ones~\cite{us}. 
However, their effect is neglected within the effective
Hamiltonian of Eq.~(\ref{eq:effective_Hamiltonian}).

\section{Conclusions}\label{conclusions}
 
We have developed a PPS scheme to study the time-resolved 
dynamics of fermionic and bosonic impurities immersed in a harmonically confined BEC.
Coherence properties and induced-interactions are encoded in the probe spectra for both 
attractive and repulsive interactions. 
Moreover, long-lived attractive and repulsive polarons exist up to $g_{BI}\approx g_{BB}$. 
For $g_{BI}>g_{BB}$, with the dynamics being dominated by energy redistribution processes, 
a rather rapid temporal orthogonality catastrophe occurs. 
To explicitly showcase that energy redistribution processes take place we have discussed the behavior of the  
corresponding interspecies interaction energy which decreases for short-evolution times and thus captures the energy transfer 
from the impurities to the environment. 
Furthermore, it shows a saturation tendency for large evolution times, a behavior that is indicative of a relaxation process of the 
impurities. 
Indeed, at longer times ($t_d>100\omega^{-1}$), where any coherence information is lost, a thermalized state is reached. 
To further characterize the aforementioned relaxation dynamics we have resorted to an effective ETH model which has enabled us 
to identify that for strong interspecies couplings and long evolution times the impurities 
acquire an effective temperature. 
This effective temperature is found to be smaller for the fermionic impurities than the bosonic ones. 
Importantly, we have found that the thermalization process is independent of the size of the bath and the impurity 
concentration, the interacting nature of the impurities as well as their flavor 
and mass. 

It would be intriguing to utilize PPS at finite temperature~\cite{Levinsen_T,Meera_T} 
and also in higher dimensions to explore the time-resolved formation of quasiparticles. 
As further perspectives, PPS could be exploited to unravel re-condensation dynamics \cite{Kock,Luo} in 
excited-bands of optical lattices and the dynamics of vibrational states of ultra-long range 
Rydberg molecules \cite{Rydberg,Rydberg1} to infer their lifetime.

\appendix 

\section {Details of the Reverse Radiofrequency spectroscopy} \label{rf_implement}

Let us elaborate on the model that allows for the
simulation of the MB dynamics under the influence of radiofrequency fields \cite{Kohstall,Scazza}.
This model has been employed in the main text for the characterization of the coherence
properties of polaronic quasiparticles in the context of pump-probe spectroscopy (PPS).

In our case few atomic impurities are immersed in a BEC environment close to an interspecies
magnetic Feshbach resonance~\cite{Chin}. 
The case of bosonic impurities possessing equal
mass to the BEC atoms can be realized by employing different hyperfine
states of a particular isotope e.g. ${}^{85}$Rb or ${}^{87}$Rb. 
For fermionic impurities the equal mass scenario  
occurs approximately e.g. for ${}^{173}$Yb impurities immersed
in a ${}^{174}$Yb BEC with mass ratio of $m_B/m_I \approx 1.006$. 
Different masses for the impurities and the BEC atoms can be realized by invoking different atomic 
species, e.g. considering ${}^{87}$Rb and ${}^{133}$Cs~\cite{widera}. 
Typically, atoms close to a broad Feshbach resonance experience a sizeable quadratic Zeeman shift
or are close to the Paschen-Back regime \cite{Chin}. 
For instance Feshbach resonances occur at magnetic fields of the order 
of $800$ G for ${}^{85}$Rb atoms where for comparison the Paschen-Bach
regime for the ground state $5~{}^{2}{\rm S}_{1/2}$ occurs for $B>2000$ G \cite{Steck}. 
This sizeable Zeeman shift allows us to address selectively the distinct $m_F$ transitions provided 
that the intensity of the radiofrequency pulse results in a Rabi frequency $\Omega_R$ much 
smaller than the Zeeman splitting of the involved hyperfine levels. 
The latter, is typically of the order of a few tenths of MHz. 
\begin{figure}[h]
    \centering
    \includegraphics[width=1.0\columnwidth]{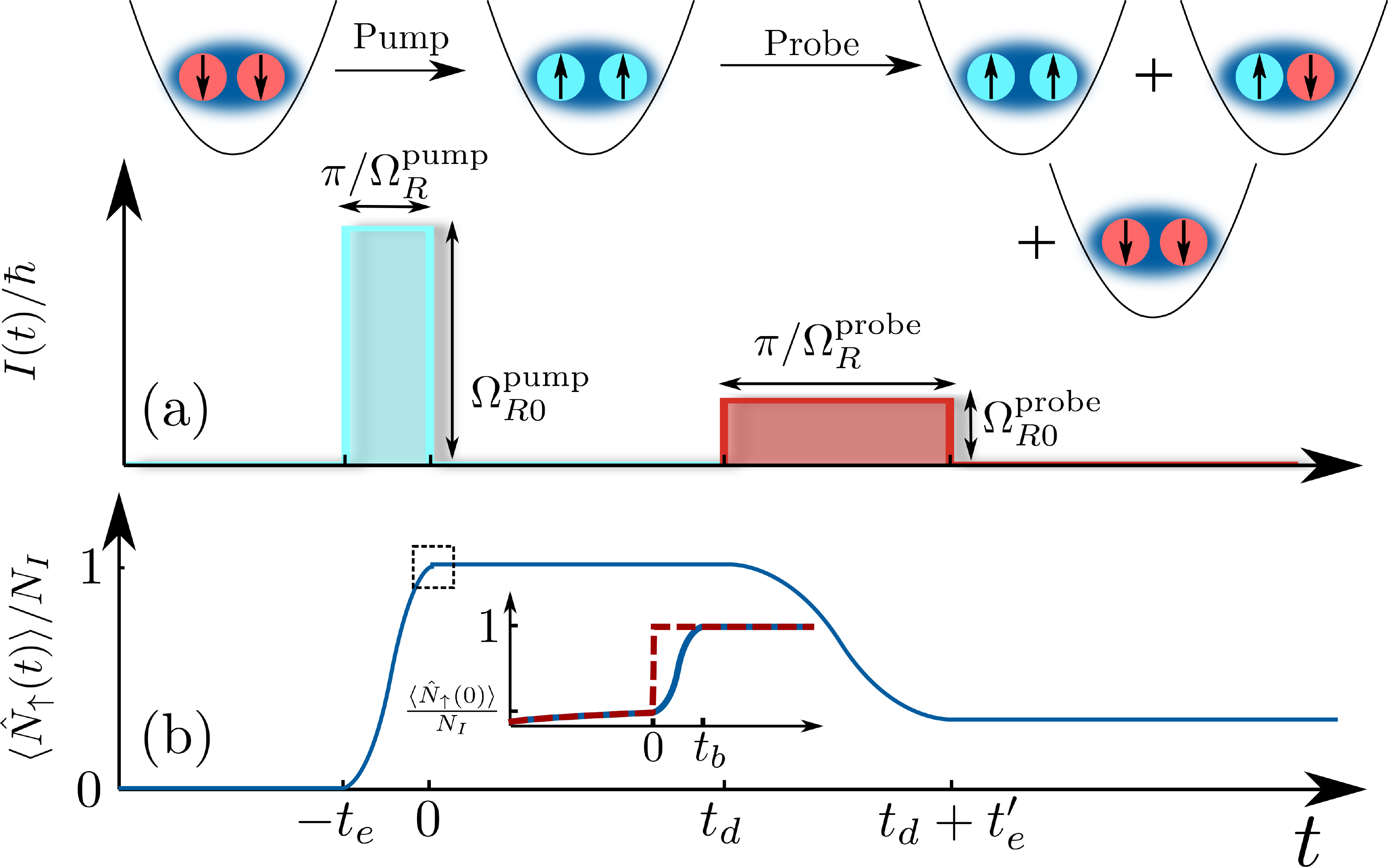}
    \caption{ (a) Schematic illustration of the employed PPS pulse sequences. 
    The involved spin configuration at each state of the dynamics is also provided. 
    (b) Expected time-evolution of 
    the population of spin-$\uparrow$ atoms during the PPS sequence. The inset depicts the 
    evolution of $\langle \hat{N}_{\uparrow}(t) \rangle/N_I$ during the application of the optical 
    burst pulse (blue line) and the approximation of employing the projection operator $\hat{P}$ at 
    $t=0$ (dark red line).}
    \label{fig:schematic}
\end{figure}
This large splitting of the different $m_F$ levels implies that magnetic phenomena such as spin-exchange 
interactions can be safely neglected for these values of the magnetic field, see also Appendix~\ref{dimensional_reduction}. 

In this work we consider two hyperfine levels of the impurity atoms 
denoted as $\ket{\uparrow}$ and $\ket{\downarrow}$. 
These states can be identified and resonantly coupled 
for a frequency $\nu_{0}$, corresponding to the
Zeeman splitting between the two levels, when a BEC environment is absent. 
Due to the harmonic confinement of the atoms each of the hyperfine levels is further divided into 
states of different atomic motion.
The average spacing between these sublevels is of the order of the harmonic trap frequency,
$\hbar \omega$ lying within the range of a few tenths of $h \times$Hz to a few $h \times$kHz in typical ultracold atom 
experiments~\cite{Nobeler,Engels}. 
In the vicinity of a Feshbach resonance the energy of these
sublevels strongly depends on the interspecies interaction strength, $g_{BI}$, 
between the impurity atoms in the resonantly-interacting hyperfine state, 
and their BEC environment. 
Accordingly the energy of each motional state shifts by $\Delta_{+}(g_{BI})$, from the 
corresponding non-interacting one. 
As it is also made obvious within the main text [see Fig.~\ref{Fig:1}(b)] this shift is of 
the order of $\omega$ to $\sim 10\omega$. 
Therefore, due to the separation of the different involved energy scales it suffices to treat the
impurities as two-level atoms. 
Furthermore, even for $\Omega_{R 0}^{\beta} \gtrapprox \Delta_{+}\sim$
kHz where the regime of strong intense pulses is accessed, $\Omega_{R 0}^{\beta}
\ll \nu_{0} \sim 10$ MHz, allowing us to invoke the rotating wave
approximation. 
Notice here that $\beta \in \left\{ \rm{pump}, \rm{probe}, \rm{dark} \right \}$. 
Within this approximation and in the interaction picture of the 
$\ket{\uparrow} \leftrightarrow \ket{\downarrow}$
transition, the Hamiltonian for the internal state of the impurities reads
$\hat{H}_S=-\frac{\hbar \Delta^{\beta}}{2} \hat{S}_z+\frac{\hbar \Omega_{R 0}^{\beta}}{2} \hat{S}_x$, which is
exactly the form employed in Eq.~(\ref{Htot_system}) of the main text. 
$\Omega_{R 0}^{\beta}$ and $\Delta^{\beta}=\nu^{\beta} - \nu_0^{\beta}$ refer respectively to the (bare) Rabi 
frequency and the detuning with respect to the resonance of the 
$\ket{\uparrow} \leftrightarrow \ket{\downarrow}$ transition at $g_{BI}=0$. 
We remark that the $\ket{\uparrow}$ and $\ket{\downarrow}$ states in 
the Schr{\" o}dinger and interaction pictures are equivalent, 
so our conclusions are invariant under this frame transformation~\cite{ferro_kout}. 

To populate the polaronic states we employ a pump pulse of rectangular 
shape as depicted in Fig.~\ref{fig:schematic}(a). 
The system is initialized in the non-interacting ground state where the impurity atoms are
spin-polarized in their $\ket{\downarrow}$ state. 
The pump pulse is characterized by frequency $\nu^{\rm pump}$, and a detuning 
$\Delta^{\rm pump}$ is employed. 
This pulse is further characterized by an exposure time $t_{e}$ 
and a bare Rabi-frequency $\Omega^{\rm pump}_{R 0}$. 
Different realizations utilize different detunings $\Delta^{\rm pump}$ and exposure times 
$t_{e}$ but the same $\Omega^{\rm pump}_{R 0}$. 
In the duration of the pulse the system undergoes Rabi-oscillations 
which for strong enough pulses $\Omega^{\rm pump}_{R 0} \gg\omega$ 
are well characterized by a Rabi-frequency 
$\Omega_{R}(\Delta^{\rm pump})=\sqrt{(\Omega_{R+}^{\rm pump})^2+(\Delta^{\rm pump} -
\Delta_+)^2}$, where $\Omega_{R+}^{\rm pump}$ and $\Delta_+$ are the corresponding
resonance values. 
By fitting the spectroscopic signal, which is the fraction of atoms transferred to
the $\ket{\uparrow}$ hyperfine state, to the theoretical 
lineshape for rectangular pulses reading
\begin{equation}
    \frac{ \langle \hat{N}_{\uparrow} (t) \rangle }{N_I}=\left[ \frac{\Omega^{\rm pump}_{R+}}
    {\Omega^{\rm pump}_{R}(\Delta^{\rm pump})} \right]^2 \sin^2\left(  \frac{\Omega^{\rm pump}_{R}
    (\Delta^{\rm pump}) t_{e}}{2} \right),
    \label{eq:lineshape}
\end{equation}
these resonance values of $\Omega_{R+}^{\rm pump}$ and $\Delta_+$ can be obtained. 
Note here that the lineshape [Eq.~(\ref{eq:lineshape})] exhibits an infinite 
sequence of peaks at the locations, $\Delta_n^{\rm pump}$, $n=0,\pm 1,\dots$, 
given by the solutions of
\begin{equation}
\frac{\Omega^{\rm pump}_R(\Delta^{\rm pump}) t_{e}}{2} = \tan \left( \frac{\Omega^{\rm pump}
_R(\Delta^{\rm pump}) t_{e}}{2} \right)
    \label{eq:peakslocs}
\end{equation}
for $\Delta^{\rm pump}$. 
Solving numerically Eq.~(\ref{eq:peakslocs}) we can identify the
location of the three first peaks at positions $\Delta_{0}^{\rm pump}=\Delta_+$ 
and $\Delta_{\pm 1}^{\rm pump} \approx \Delta_+ \pm \Omega_{R+}^{\rm pump} \sqrt{(\frac{8.9868}
{\Omega_{R+}^{\rm pump}t_{e}})^2-1}$. 
Their corresponding amplitudes read $A_0 = \sin^2(\frac{1}{2}\Omega^{\rm pump}_{R+} t_{e})$ 
and $A_{\pm 1} \approx  0.01179 (\Omega_{R+}^{\rm pump} t_{e})^2$. 
In order to achieve a high spectroscopic signal, $\braket{\hat{N}_{\uparrow}(t_d)}/N_I$, we set the exposure time 
to $t_{e}=\pi/\Omega^{\rm pump}_{R+}$ (up to the obtained fitting accuracy) 
ensuring that $A_0 \approx 1$. 
This choice implies that the peaks at $\Delta^{\rm pump}_{\pm 1}$ are clearly imprinted in the
obtained spectrum possessing an amplitude $A_{\pm 1} \approx 0.116438$. 
Indeed, these side-peaks can be clearly identified in Fig.~\ref{Fig:1}(b).

To infer about the coherence properties of the polaronic states we employ PPS, 
see Fig.~\ref{fig:schematic}(a).
Initially, we prepare the system in the same non-interacting ground state as in the previously
examined protocol and apply a rectangular $\pi$-pulse, with $\Omega^{\rm pump}_{R0}=10\omega$ 
and $t_{e}=\pi/\Omega^{\rm pump}_{R+}$ on a polaronic resonance where we have
identified the resonant $\Omega^{\rm pump}_{R+}$ and $\Delta^{\rm pump}_+$ as explained above. 
This sequence transfers the atoms from the ground state to the polaronic state in a very efficient
manner, see Fig.~\ref{fig:schematic}(b). 
Then the impurity atoms are projected to the spin-$\uparrow$ state by employing an optical
burst transition on the lowest hyperfine state $| \downarrow \rangle$ to an available $P$ 
electronic level at $t=0$ which essentially ejects all the spin-$\downarrow$ atoms from the trap. 
This procedure has been simulated by the application of the operator 
$\hat{H}_P=-i \hbar \Gamma \int dx~\hat\Psi^{\dagger}_{\downarrow}(x) \hat\Psi_{\downarrow}(x)$ 
over a short time interval $t_b$. 
We can numerically verify that for large $\Gamma > 100 \omega$ and small $t_b \ll \omega^{-1}$ 
(corresponding to the experimentally relevant values) the action of $\hat{H}_P$ to the state
after the pump pulse is equivalent to the projection of the impurity to the spin-$\uparrow$
configuration. 
For this reason and for computational simplicity we employ the state $| \Psi (t=0^+)
\rangle= \frac{\hat{P} | \Psi(t=0^-)\rangle}{||\hat{P} | \Psi(t=0^-)\rangle||}$ 
as an initial state for the subsequent time-evolution $t>0$. 
Note that this sequence for $\Omega^{\rm pump}_R \ll\omega$ 
is approximately equivalent to an interaction quench, as the pump-pulse has almost no
spectral selectivity due to the pronounced power-broadening of the radiofrequency transition, as
$\Omega_{R}^{\rm pump} \sim \Delta^{\rm pump}_+$ and the fact that the out-of-equilibrium dynamics
is effectively frozen due to the small time-scale 
$t_{e}=\pi/\Omega^{\rm pump}_R \ll \omega^{-1}$. 
Indeed, these properties of the pump-pulse have been verified numerically for the selected 
parameters $\Omega^{\rm pump}_{R 0}$ and $\Delta^{\rm pump}$ as the MB state after 
this pulse is found to possess a fidelity in excess of $90\%$ to the initial one. 

After the pump sequence is completed we let the system evolve in the absence
of radiofrequency fields, $\Omega^{\rm dark}_{R 0}=0$, for a dark time, $t_d$. 
Finally, we apply a second probe $\pi$-pulse with a smaller 
$\Omega^{\rm probe}_{R 0}=1\omega$ to the first one and varying 
$\Delta_+^{\rm probe}$ to transfer the atoms from the polaronic to the initial ground state. 
The employed spectroscopic signal is the fraction of atoms that have been 
deexcited by the probe pulse (recall that within our scheme all of the 
particles are at $t_d$ in the spin-$\uparrow$ state) divided by the total
number of impurities, $\frac{\langle N_{\downarrow} (t_d) \rangle}{N_I}$. 
Note that a smaller $\Omega^{\rm probe}_{R 0}$, when compared 
to $\Omega^{\rm pump}_{R 0}$, is employed in order to reduce
the power-broadening during the probe sequence and subsequently
increase the resolution in terms of detuning. 
However, this value cannot be arbitrarily lowered 
since for decreasing probe intensities the frequency-resolution 
is increased at the expense of lower temporal-resolution. 
For such low intensities the motional state of the spin-$\uparrow$ impurities is significantly
altered during the application of the probe pulse. 
As a heuristic argument the relation $\delta \nu \delta t\approx 1$ 
that connects the temporal ($\delta t$) and spectral ($\delta \nu$) resolution is
commonly employed~\cite{Sakurai}. 
The value of $\Omega^{\rm probe}_{R 0} = 1\omega$ is selected within this
work as it consists an adequate tradeoff between the spectral and the temporal resolution. 
Finally, owing to the rectangular shape of the probe pulse the exhibited lineshape 
of $\frac{\langle \hat{N}_I (t_d)\rangle}{N_I}$ is given by Eq.~(\ref{eq:lineshape}) 
as long as the impurity is coherent, i.e. $| g^{(1)}(x,x';t_d)| \approx 1$. 
Accordingly, in our analysis we attribute all fringes appearing in the spectra 
to the lineshape of a single resonance if the ratio of the
amplitude of two neighboring peaks satisfies $\frac{A_{n+1}}{A_{n}}< 0.12$.

\section{Comparison with Ramsey spectroscopy}\label{Ramsey} 

Next we demonstrate the advantage of utilizing the PPS scheme in comparison to Ramsey spectroscopy.  
In particular, we explicitly showcase the differences between the predictions of the PPS and the Ramsey schemes 
for intriguing phenomena exhibited by our system including the TOC and the thermalization process. 
To achieve this comparison we have simulated the Ramsey response of our system following the scheme 
described in Refs.~\cite{Us_Busch,us}. 
The main facet of this Ramsey protocol is that by applying an intense radiofrequency pulse to the initially 
noninteracting with the bath spin-$\downarrow$ impurities we transfer them into a superposition state 
$\frac{\ket{\uparrow}+\ket{\downarrow}}{\sqrt{2}}$, where the state $\ket{\uparrow}$ interacts with the bosonic medium. 
Thus, the time-evolved MB wavefunction e.g. of a single impurity is given by 
$\ket{\Psi(t)}=(1/\sqrt{2}) e^{-i \hat{H}_Rt/ \hbar } \ket{\Psi_{BI}^0} \ket{\uparrow}+ 
(1/\sqrt{2}) e^{-i E_0 t/\hbar} \ket{\Psi_{BI}^0} \ket{\downarrow}$.  
Here, $\hat{H}_R = \hat{H}^{0}_{B}+\hat{H}_{BB}+\hat{H}^{0}_{\uparrow} +\hat{H}^{0}_{\downarrow}+\hat{H}_{BI}$ as introduced 
in Sec.~\ref{model}, $\ket{\Psi_{BI}^0}$ is the spatial part of the initial MB wavefunction [see also Appendix~\ref{sec:method}] 
and $E_0$ refers to the corresponding eigenenergy. 
In this protocol the structure factor, $|S(t)|=|\braket{\Psi_{BI}^0|e^{i E_0 t/\hbar} e^{-i \hat{H}_Rt/ \hbar }|\Psi_{BI}^0}|$, 
of the system is monitored by inspecting the magnitude of the impuritys' spin $|\braket{\hat{{\bm S}}(t)}|$. 
\begin{figure}
 	\includegraphics[width=0.48\textwidth]{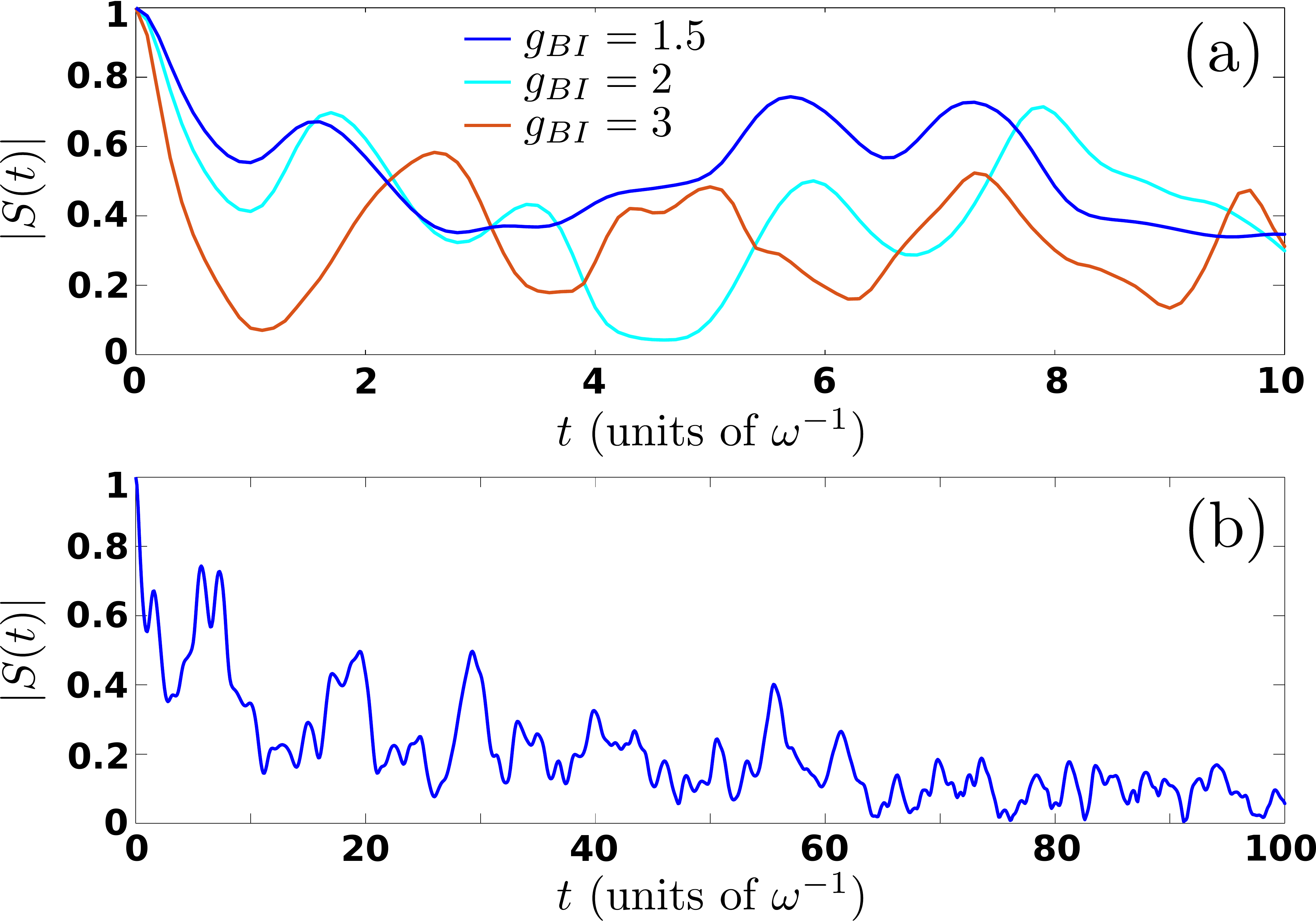}
        \caption{(a) Time-evolution of the structure factor of one impurity at different impurity-BEC 
        interaction strengths (see legend).  
        (b) Dynamics of the structure factor depicted in (a) of one impurity for 
        $g_{BI}=1.5\sqrt{\hbar^3 \omega/m}$ over a longer timescale.
        The bosonic bath contains $N_B=100$ bosons. 
        The system is initialized in its ground state with 
        $g_{BI}=0$.}
 	\label{fig:structure_factor_dif} 
\end{figure}

According to our discussion in Sec.~\ref{PPS_details} it becomes apparent that time-dependent 
phenomena such as the TOC and the consecutive thermalization of the impurities can be clearly tracked in an experimentally relevant fashion 
via employing the temporarily-resolved PPS scheme. 
Indeed, Figs.~\ref{Fig:1}(i)-(k) in the main text reveal that TOC takes place already for 
$t_d=2\omega^{-1}$, 
with the presence of quasi-free impurities at $\Delta^{{\rm probe}}\approx 0$ being also readily imprinted in the probe spectrum. 
On the contrary, the only information that Ramsey spectroscopy conveys is the 
value of the structure factor, $\abs{S(t)}$. 
Importantly, it does not deliver any further insights 
about the physical origin of its decreasing tendency and thus the underlying physical processes, 
see in particular Figs.~\ref{fig:structure_factor_dif}(a), (b). 
Indeed, for $t<10\omega^{-1}$ $\abs{S(t)}$ oscillates having a minimum value of $0.4$ when 
$g_{BI}=1.5\sqrt{\hbar^3 \omega/m}$ and thus does not provide any clear 
signature for the emergence of the TOC identified using PPS. 
Along the same lines for $g_{BI}=1.5\sqrt{\hbar^3 \omega/m}>g_{BB}$, namely after the TOC manifests itself, a 
thermalization tendency is clearly imprinted in the probe spectrum [see Figs.~3(g)-(i)] with a predominant peak appearing
at $\Delta^{{\rm probe}}\approx 0$. 
In other words, the dynamics after the
decay of the strongly ($g_{BI}>g_{BB}$) repulsive Bose polarons leads to a quasi-stationary 
state of the MB system with respect to the energy redistribution among the different dynamical modes, 
providing this way strong evidences towards a thermalized state. 
This mechanism cannot be even suggested by invoking the contrast computed within the 
above-discussed Ramsey scheme. 
Indeed, the Ramsey scheme only indicates the tendency towards thermalization due to the decreasing behavior 
of the structure factor which however fluctuates within the depicted time-interval.

\section{Dimensional reduction of the many-body Hamiltonian from three- to one-dimensions} \label{dimensional_reduction}

We consider an ensemble of confined ultracold atoms in three different hyperfine
states, denoted as $B$, $\uparrow$ and $\downarrow$. 
State $B$ is occupied by bosonic bath particles and the $\uparrow$ and
$\downarrow$ states constitute a pseudo-spin-$1/2$ sub-system. 
We assume that state $B$ belongs to a different manifold of hyperfine states with respect to
the total angular momentum quantum number, $F$, than the other two pseudospin states. 
The system is optically trapped and therefore all of the above hyperfine states experience 
the same confinement. 

The {\it ab-initio} Hamiltonian of this multicomponent system reads 
$\hat{H}=\hat{H}_0+\hat{H}_{\rm SD}+\hat{H}_{\rm I}$. 
The spin-independent part $\hat{H}_0$ is given by
\begin{equation}
    \hat{H}_{\rm 0} = \int {\rm d}^3 r \sum_{\sigma \in \{ B, \uparrow,\downarrow \}}
    \hat{\psi}_{\sigma}^{\dagger}({\bm r}) \left( -\tfrac{\hbar^2}{2 m} \nabla^2 + V_0({\bm r})
    \right) \hat{\psi}_{\sigma}({\bm r}),
    \label{eq:sp_ham3}
\end{equation}
where $m$ is the mass of the chemical element and $V_0({\bm r})$ refers to the confining
potential. 

By imposing a homogeneous magnetic field, along the $z$-direction, the energy of the magnetic sublevels characterized by
different $m_F$ shift due to the Zeeman effect. 
Accordingly, the state-dependent part of the Hamiltonian is expressed as
\begin{equation}
    \begin{split}
    \hat{H}_{\rm SD}=& E_B \hat{N}_{B}
        +\frac{E_{\uparrow}+E_{\downarrow}}{2} \hat{N}_I \\
        &+\frac{E_{\uparrow}-E_{\downarrow}}{2}
        \sum_{\alpha,\beta \in \{ \uparrow,\downarrow \}} \int {\rm d}^3 r~ \hat{\psi}_{\alpha}^{\dagger}({\bm r})   \sigma^{z}_{\alpha \beta}
\hat{\psi}_{\beta}({\bm r}).
    \end{split}
    \label{eq:sd_ham3}
\end{equation}
where $\sigma^{z}_{\alpha \beta}$ corresponds to the spin-$z$ Pauli matrix and $E_{\sigma}$ is the
energy of the atomic state of species $\sigma \in \{B,\uparrow,\downarrow\}$. 
The typical energy difference between hyperfine levels possessing different $F$ is of the order of
several $h \times$GHz. 
In particular, for ${}^{87}$Rb the hyperfine splitting between the two lowest hyperfine
manifolds $F=1$ and $F=2$ is $E_{F=2}-E_{F=1} \approx h \times 6.83$ GHz \cite{Steck}. 
Additionally, the amplitude of the Zeeman energy shifts is also of the order of $h \times$ MHz/G. 
For instance in ${}^{87}$Rb this amplitude is of the order of $\sim 0.7~h \times$ MHz/G \cite{Steck}. 
Furthermore, in the same species quadratic Zeeman shifts that lead
to a non-equidistant distribution of magnetic sublevels possessing an amplitude of several $h
\times$ MHz can be observed
already for magnetic fields of the order of $\sim 10$ G \cite{Steck}. 
Typical ultracold atom experiments involve interaction energies ranging from hundreds of $h \times$ Hz to a few $h \times$ kHz generating this way interaction energy 
shifts and spin-exchange processes characterized by energies of the same order of magnitude. Therefore, except for the case that the magnetic field applied is of the order
of few Gauss the spin-exchanging collisions are strongly suppressed \cite{Bersano}. 

Operating in the ultracold limit dominated by $s$-wave scattering \cite{Pethick} the interaction 
Hamiltonian can be expressed as \cite{Huang}
\begin{equation}
\hat{H}_{\rm I} = \sum_{\sigma,\sigma'} \tfrac{4 \pi \hbar^2
a_{\sigma \sigma'}}{m} \int {\rm d}^3 r~ \hat{\psi}_{\sigma}^{\dagger}({\bm
    r})\hat{\psi}_{\sigma'}^{\dagger}({\bm r}) \hat{\psi}_{\sigma'}({\bm
    r})\hat{\psi}_{\sigma}({\bm r}).
    \label{eq:int_ham3}
\end{equation}
The scattering lengths $a_{\sigma \sigma'}$, with $\sigma' \in \{ B, \uparrow, \downarrow \}$, can be tuned via a Fano-Feshbach 
resonance between two distinct hyperfine levels \cite{Chin}. 

In order to effectively reduce the dimensionality of the above system from three- (3D) to one-dimension (1D) a strong confinement 
along the two perpendicular spatial directions is usually employed \cite{Strigari}. 
Then, the confining potential reads
\begin{equation}
    V_0({\bm r})= \frac{1}{2} m \omega^2 x^2 + \frac{1}{2} m \omega^2_{\perp} (y^2+z^2),
    \label{eq:cp_3t1}
\end{equation}
where $\omega_{\perp}\gg \omega$ holds for the transverse and longitudinal trapping frequencies.
Note that the potential of Eq. (\ref{eq:cp_3t1}) can be realized either by a single optical dipole trap \cite{Bersano}, or by
applying a deep two-dimensional optical lattice potential \cite{Catani,Negerl}. 
To access the 1D regime, the frequency of the transverse confinement $\omega_{\perp}$ has to be selected such that the
excited states of the harmonic trap along the transverse directions ($y$, $z$)  are not populated. 
The condition for a 1D BEC is well-known \cite{Strigari} and reads $N_{B} a_{BB} \alpha_{\perp}/\alpha^2 \ll 1$,
where $\alpha_{\perp}=\sqrt{\hbar/m \omega_{\perp}}$ and $\alpha=\sqrt{\hbar/m \omega}$.
In the few atom case, referring to the impurity species, a sufficient condition for accessing the 1D limit
is $\omega_\perp \gg N \omega$ \cite{Idziaszek}. 
Indeed, under this assumption it is known that even in the strong interaction
limit the system behaves as a Tonks-Girardeau gas of hard-core bosons sharing some characteristics with a gas of free 
fermions of the same particle number \cite{Girardeau,Lieb}. 
Properties of such fermionized 1D bosons have been probed experimentally in Refs. \cite{BlochTonks,WeissTonks}. 

Accordingly, the corresponding 3D field operators can be expressed in terms of 1D ones as follows 
\begin{equation}
    \hat{\psi}_{\sigma}^{\dagger}({\bm r})=\sqrt{\frac{m \omega_{\perp}}{\pi \hbar}} e^{-\frac{m
    \omega_{\perp}}{2 \hbar}(y^2+z^2)} \hat{\psi}_{\sigma}^{\dagger}(x).
    \label{eq:fo_3t1}
\end{equation}
By employing Eq. (\ref{eq:fo_3t1}) we can then evaluate straightforwardly the reduced 1D effective 
Hamiltonian for $\hat{H}_{\rm 0}+\hat{H}_{\rm
SD}$ which takes the form
\begin{equation}
    \begin{split}
    &\hat{H}_{\rm 0} + \hat{H}_{\rm SD}= \frac{E_{\uparrow}-
    E_{\downarrow}}{\hbar} \hat{S}_z+ \\&\sum_{\sigma \in \{B, \uparrow,\downarrow \}} \int {\rm d} x~ 
    \hat{\psi}_{\sigma}^{\dagger}(x) \left( -\tfrac{\hbar^2}{2 m} \tfrac{{\rm d}^2}{{\rm d}x^2} +
    \tfrac{1}{2} m \omega^2 x^2 \right) \hat{\psi}_{\sigma}(x). 
    \end{split}
    \label{eq:sp_ham1}
\end{equation}
Here, all terms contributing to the total energy shift for constant $N_I$ and $N_B$ are dropped while the $\hat{S}_z$ operator reads 
\begin{equation}
    \hat{S}_z = \frac{\hbar}{2} \int {\rm d} x
    ~\left( \hat{\psi}^{\dagger}_{\uparrow}(x)\hat{\psi}_{\uparrow}(x)
    -\hat{\psi}^{\dagger}_{\downarrow}(x)\hat{\psi}_{\downarrow}(x) \right).
    \label{eq:sd_ham1}
\end{equation} 
The dimensional reduction of $\hat{H}_I$ is, however, more complicated. 
In particular, the phenomenon of the confinement induced resonance \cite{Olshanii,Negerl2} occurs when 
$\alpha_{\perp}=\sqrt{\frac{\hbar}{m \omega_{\perp}}}$ 
is comparable to $a_{\sigma \sigma'}$. 
This implies that the actual 1D coupling constant deviates from $g_{\sigma \sigma'}^{\rm
MF}=\frac{2 \hbar^2 a_{\sigma \sigma'}}{m a_\perp^2}$ \cite{Strigari} which is obtained by evaluating the integrals
appearing in Eq. (\ref{eq:int_ham3}) along the transverse ($y$, $z$) directions. 
Detailed theoretical and experimental investigations \cite{Olshanii,Negerl2} reveal
that the 1D coupling strength $g_{\sigma \sigma'}$ possesses a simple analytical form
i.e. $g_{\sigma \sigma'}=g_{\sigma \sigma'}^{\rm MF} (1-\frac{|\zeta(1/2)| a_{\sigma \sigma'}}{\sqrt{2} a_{\perp}})^{-1}$, and 
the effective 1D interaction Hamiltonian simplifies to
\begin{equation}
    \hat{H}_{\rm I} = \sum_{\sigma,\sigma'} g_{\sigma \sigma'} \int {\rm d} x ~ \hat{\psi}_{\sigma}^{\dagger}(x)
    \hat{\psi}_{\sigma'}^{\dagger}(x)\hat{\psi}_{\sigma'}(x)\hat{\psi}_{\sigma}(x).
    \label{eq:int_ham1}
\end{equation} 
Note here that due to the double counting for intraspecies interaction terms in Eq.~(\ref{eq:int_ham1}), 
the parameter $g_{BB}$ appearing in the latter is two times larger than the corresponding 
one that is involved in Eq.~(\ref{Htot_system}).

According to the above discussion, for the experimental implementation of the setup descibed in the main text the 
interaction parameters, $g_{BB}$, $g_{BI}$ and $g_{II}$, used 
herein are related to the corresponding 3D scattering lengths as follows
\begin{equation}
    a_{\sigma \sigma'}=\alpha_{\perp} \frac{2 \tilde{g}_{\sigma \sigma'}}{\sqrt{2} |\zeta(1/2)|
        \tilde{g}_{\sigma \sigma'} + 8 \eta}.
    \label{eq:scat_length}
\end{equation}
Here, $\tilde{g}_{\sigma \sigma'}=g_{\sigma \sigma'}/\sqrt{\hbar^3 \omega/m}$ refers to the
dimensionless interaction strength and $\eta= \alpha/\alpha_{\perp} = \sqrt{ \omega_{\perp}/
\omega}$ is the aspect ratio.
Furthermore, BECs involving particle numbers of the order of $N_B \sim 100$ are already accsessible by current 
state-of-the-art experimental settings e.g. in optical lattice experiments \cite{Catani,Negerl}. 
Finally, it is worth commenting that three-body recombination processes are highly suppressed for alkali ultracold 
atomic vapors, as the one considered herein. 
For instance, for a ${}^{87}$Rb BEC and in the presence of three-body recombination a lifetime of $14.8$s has been 
reported \cite{Dalibard}. 
Note also that the rate of three-body recombination scales with the cube of the density and as a consequence,
this effect is negligible for the mesoscopic system under consideration which involves low densities.

\section{The many-body variational methodology: ML-MCTDHX} \label{sec:method}

To track the stationary properties and most importantly the MB quantum dynamics of the multicomponent system addressed in the main text we resort to 
the Multi-Layer Multi-Configuration Time-Dependent Hartree method for atomic mixtures (ML-MCTDHX) \cite{MLB1,MLX,MLB2}. 
It constitutes an ab-initio variational method for solving the time-dependent MB Schr{\"o}dinger equation of atomic mixtures possessing 
either bosonic \cite{us_phase_sep,darkbright,Mistakidis_bose_pol} or fermionic \cite{ferro_kout,Us_Fermi,phase_sep_ferm,Siegl} spinor components. 
A major advantage of this approach is the expansion of the total MB wavefunction with respect to a time-dependent and variationally 
optimized basis (see below). 
This allows us to capture all the relevant inter- and intraspecies correlations of a multicomponent system in an efficient manner 
at each time instant by utilizing a reduced number of basis states when compared to expansions relying on a time-independent basis. 

The system considered in the main text consists of a bosonic bath (B) with $N_B=100$ atoms and either one ($N_I=1$) or two ($N_I=2$) 
impurity (I) atoms. 
Most importantly, the impurities being either bosons or fermions possess an internal pseudospin-$1/2$ degree of freedom \cite{Kasamatsu,Us_Busch}. 
To account for interspecies correlations, the MB wavefunction $|\Psi(t)\rangle$ is expressed according to a truncated Schmidt 
decomposition \cite{Horodeckix4,darkbright,us_phase_sep} in terms of $D$ different species functions, i.e. $|\Psi^{\sigma}_k(t)\rangle$, 
for each component $\sigma=B,I$. 
We remark that the time-dependent species functions $|\Psi^{\sigma}_k(t)\rangle$ form an orthonormal $N_{\sigma}$-body wavefunction set 
within a subspace of the $\sigma$-species Hilbert space $\mathcal{H}^{\sigma}$ \cite{MLX}. 
Then, the MB wavefunction $\ket{\Psi(t)}$ ansatz reads   
\begin{equation} 
    |\Psi(t)\rangle=\sum_{k=1}^D \sqrt{\lambda_k(t)} |\Psi^{\rm B}_k(t)\rangle|\Psi^{\rm I}_k(t)\rangle,   
    \label{eq:wfn}
\end{equation} 
where the time-dependent Schmidt weights $\lambda_k(t)$ are also known as the natural species populations of 
the $k$-th species function and provide information about the degree of entanglement between the individual subsystems. 
For instance, if two different $\lambda_k(t)$ are nonzero then $\ket{\Psi(t)}$ is a linear superposition of two 
states and therefore the system is entangled \cite{Roncaglia,Horodeckix4} or interspecies correlated. 
On the other hand, in the case of $\lambda_1(t)=1$, $\lambda_{k>1}(t)=0$, the wavefunction is a direct 
product of two states and the system is non-entangled.  

Next, in order to include intraspecies correlations into our MB wavefunction ansatz each species function $\ket{\Psi^{\rm \sigma}_k(t)}$ 
is further expanded on a time-dependent number-state basis set $|\vec{n} (t) \rangle^{\sigma}$. 
Namely 
\begin{equation}
    | \Psi_k^{\sigma} (t) \rangle =\sum_{\vec{n}} A^{\sigma}_{k;\vec{n}}(t) | \vec{n} (t) \rangle^{\sigma},    
    \label{eq:number_states}
\end{equation}  
where $A^{\sigma}_{k;\vec{n}}(t)$ denote the underlying time-dependent expansion coefficients. 
Moreover, each number state $|\vec{n} (t) \rangle^{\sigma}$ corresponds to a permanent for bosons 
or a determinant for fermions building upon $d^{\sigma}$ time-dependent variationally optimized 
single-particle functions (SPFs), i.e. 
$\left|\phi_l^{\sigma} (t) \right\rangle$, with $l=1,2,\dots,d^{\sigma}$, being characterized by 
occupation numbers $\vec{n}=(n_1,\dots,n_{d^{\sigma}})$. 
Additionally, the SPFs are expanded with respect to a time-independent primitive basis. 
For the majority species, this primitive basis corresponds to an $\mathcal{M}$ dimensional discrete 
variable representation denoted in the following by $\lbrace \left| q \right\rangle \rbrace$. 
However, for the impurities the primitive basis refers to the tensor 
product $\lbrace \left| q,s \right\rangle \rbrace$ 
of the discrete variable representation basis regarding the 
spatial degrees of freedom and the two-dimensional pseudospin-$1/2$ 
basis $\{\ket{\uparrow}, \ket{\downarrow}\}$. 
Consequently, each SPF of the impurities acquires the following spinor wavefunction form  
\begin{figure}[ht]
\includegraphics[width=1.0\columnwidth]{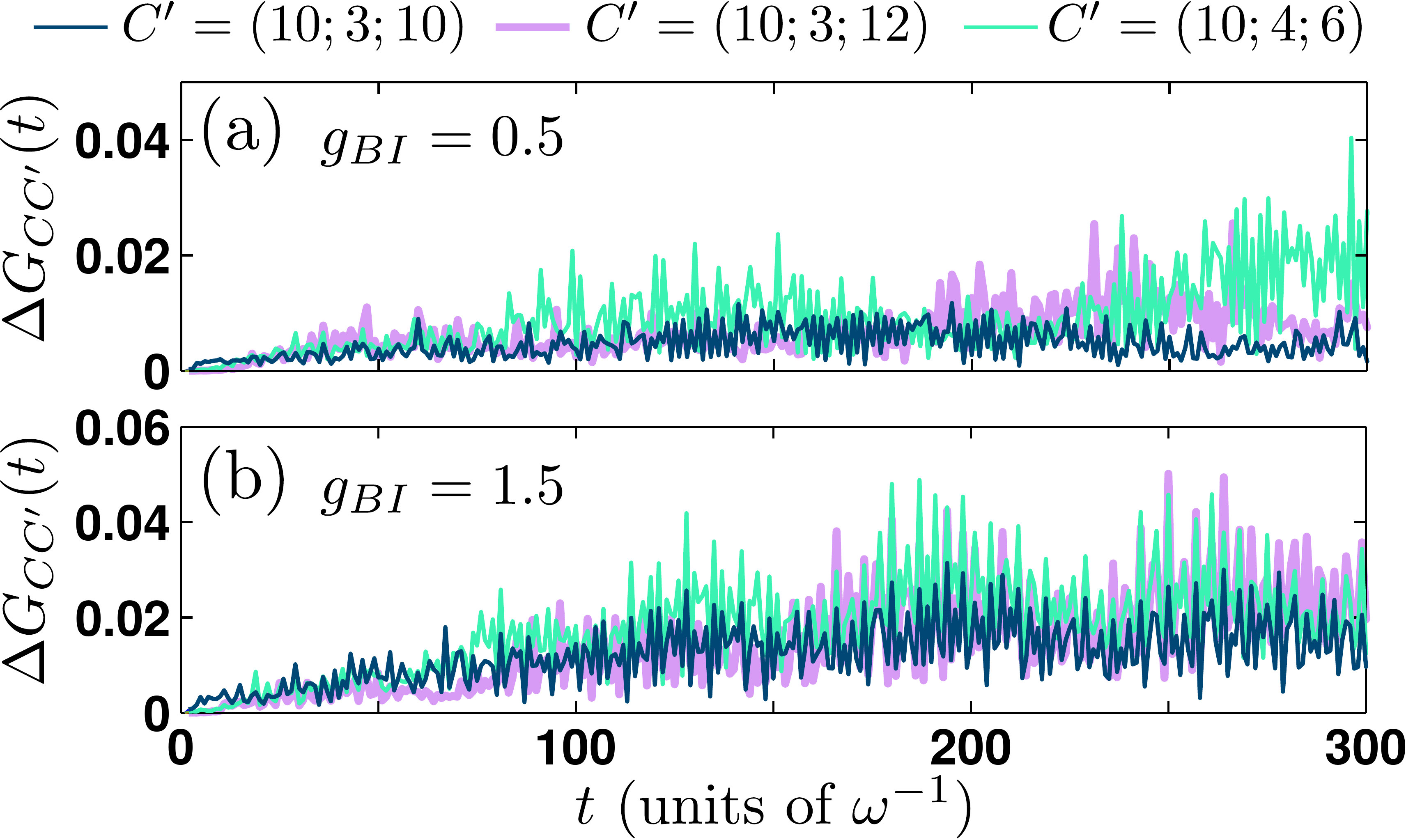}
\caption{Evolution of the one-body coherence absolute deviation $\Delta G_{C,C'}(t)$ between 
the $C=(10;3;8)$ and other orbital configurations $C'=(D;d^B;d^I)$ (see legend) 
for (a) $g_{BI}=0.5\sqrt{\hbar^3 \omega/m}$ and (b) $g_{BI}=1.5\sqrt{\hbar^3 \omega/m}$. 
In all cases $N_{B}=100$, $N_I=2$ with $g_{BB}=0.5\sqrt{\hbar^3 \omega/m}$ and $g_{II}=0$ while initially $g_{BI}=0$.} 
\label{Fig:convergence_coherence}
\end{figure}
\begin{equation}
    | \phi^{\rm I}_j (t) \rangle= \sum_{q=1}^{\mathcal{M}}\big( B^{{\rm
    I}}_{jq \uparrow}(t) \ket{q} \ket{\uparrow}+B^{{\rm I}}_{jq \downarrow}(t) \ket{q} 
    \ket{\downarrow}\big). 
    \label{eq:spfs}
\end{equation}
Here, $B^{{\rm I}}_{j q \uparrow}(t)$ [$B^{{\rm I}}_{j q \downarrow}(t)$] are the time-dependent 
expansion coefficients of the pseudospin-$\uparrow$ and $\downarrow$ respectively, 
see also Refs.~\cite{Us_Busch,ferro_kout}. 

Having exemplified the MB wavefunction ansatz and in order to address the time-evolution of the 
($N_B+N_I$)-body wavefunction $\left|\Psi(t) \right\rangle$ 
obeying the Hamiltonian of Eq.~(\ref{Htot_system}) provided in the main text we then numerically solve the so-called 
ML-MCTDHX equations of motion \cite{MLX}. 
These equations are determined by following the Dirac-Frenkel~\cite{Frenkel,Dirac} variational 
principle for the generalized ansatz 
of Eqs.~(\ref{eq:wfn}), (\ref{eq:number_states}) and (\ref{eq:spfs}). 
In this way, we obtain a set of $D^2$ linear differential equations 
of motion for the $\lambda_k(t)$ coefficients coupled to 
$D(\frac{(N_B+d^B-1)!}{N_B!(d^B-1)!}+\frac{(N_I+d^I-1)!}{N_I!(d^I-1)!})$ 
nonlinear integrodifferential equations for the species functions and $d^B+d^I$ nonlinear 
integrodifferential equations for the SPFs.

\section{Convergence of the many-body simulations}\label{convergence}  

The Hilbert space truncation within the ML-MCTDHX method is determined by the considered orbital 
configuration space i.e. $C=(D;d^B;d^I)$. 
In this notation, $D=D^B=D^I$ and $d^B$, $d^I$ denote the number of species functions and SPFs 
respectively for each species [Eqs. (\ref{eq:wfn}) and (\ref{eq:number_states})]. 
Moreover, within our numerical calculations we employ a primitive basis based on a sine discrete 
variable representation for the spatial part of the SPFs with $\mathcal{M}=600$ grid points. 
This sine discrete variable representation intrinsically introduces hard-wall boundary conditions 
at both edges of the numerical grid which in our case are located at $x_\pm=\pm50\sqrt{\hbar/m\omega}$. 
We assured that the location of the hard-wall boundaries does not impact our findings since no 
significant density portion occurs beyond $x_{\pm}=\pm20\sqrt{\hbar/m\omega}$. 
The eigenstates of the multicomponent system are obtained by utilizing the so-called improved 
relaxation method~\cite{MLX,MLB1,MLB2} 
within ML-MCTDHX. 
To address the corresponding nonequilibrium dynamics, we numerically solve the ML-MCTDHX equations 
of motion using the MB wavefunction [Eq.~(\ref{eq:wfn})] 
under the influence of the Hamiltonian (1) of the main text. 

To testify the convergence of the MB results we ensured that all observables of interest are to a 
certain level of accuracy insensitive for a varying orbital configuration space, $C=(D;d^B;d^I)$. 
Note that for the MB simulations discussed in the main text we relied on the orbital configuration 
$C=(10;3;8)$. 
To infer the convergence of our results we exemplarily showcase below the behavior of the spatially 
integrated one-body coherence function, $g^{(1)}(x,x';t)$, for different number of species 
and SPFs in the course of time. 
In particular we calculate its normalized absolute deviation between the $C=(10;3;8)$ and other 
orbital configurations $C'=(D;d^B;d^I)$, namely 
\begin{equation}
\Delta G_{C,C'}(t) =\frac{\int dx dx'\abs{g^{(1)}_C(x,x';t) -g^{(1)}_{C'}(x,x';t)}}{\int dx dx' 
g^{(1)}_C(x,x';t)}. 
\label{deviation_coherence} 
\end{equation} 
The dynamics of $\Delta G_{C,C'}(t)$ is illustrated 
in Fig.~\ref{Fig:convergence_coherence} for the 
multicomponent bosonic system consisting of $N_B=100$ atoms 
and $N_I=2$ non-interacting impurities upon considering the pump spectroscopic sequence introduced 
in Sec.~\ref{rf_implement} from $g_{BI}=0$ either 
to $g_{BI}=0.5\sqrt{\hbar^3 \omega/m}$ [Fig.~\ref{Fig:convergence_coherence}(a)] or towards 
$g_{BI}=1.5\sqrt{\hbar^3 \omega/m}$ [Fig.~\ref{Fig:convergence_coherence}(b)]. 
Evidently, a systematic convergence of $\Delta G_{C,C'}(t)$ is achieved in both cases. 
Indeed, closely inspecting $\Delta G_{C,C'}(t)$ for $g_{BI}=0.5\sqrt{\hbar^3 \omega/m}$ we observe that the deviation 
between the $C=(10;3;8)$ and $C'=(10;4;6)$ [$C'=(10;3;10)$] orbital configurations 
remains below $3.8\%$ [$1.2\%$] in the entire time-evolution 
[Fig.~\ref{Fig:convergence_coherence}(a)]. 
On the other hand for increasing $g_{BI}$, $\Delta G_{C,C'}(t)$ takes larger 
values as shown in Fig.~\ref{Fig:convergence_coherence}(b). 
For instance, at $g_{BI}=1.5\sqrt{\hbar^3 \omega/m}$ the relative error $\Delta G_{C,C'}(t)$ with $C=(10;3;8)$ and 
$C'=(10;4;6)$ [$C'=(10;3;10)$] becomes at most of the 
order of $5.2\%$ [$3\%$] at long evolution times $t>150\omega^{-1}$. 
It is also worth mentioning at this point that for all other observables and interspecies 
interaction strengths discussed in the main 
text a similar degree of convergence takes place (results not shown here for brevity).

\vspace{0.1cm}
\begin{acknowledgements} 
P.S. and G.M.K gratefully acknowledge funding by the Cluster of Excellence ``CUI: Advanced Imaging of Matter" of the Deutsche 
Forschungsgemeinschaft (DFG) - EXC 2056 - project ID 390715994. 
G.C.K. gratefully acknowledges financial support by the DFG 
in the framework of the SFB 925 ``Light induced
dynamics and control of correlated quantum systems".
S.I.M. gratefully acknowledges financial support in the framework of the Lenz-Ising Award 
of the University of Hamburg. 
T.B. has been supported by the Okinawa Institute of Science and Technology Graduate University.

\vspace{0.1cm}
G.C.K, S.I.M. and G.M.K. contributed equally to this work.
\end{acknowledgements}

{}
\end{document}